\documentclass[final,3p,times,twocolumn]{elsarticle}

\usepackage{cite}
\usepackage{amsmath,amssymb,amsfonts}
\usepackage{algorithmic}
\usepackage{graphicx}
\usepackage{textcomp}
\usepackage{bm}
\usepackage{algorithm}
\usepackage{array}
\usepackage[caption=false,font=normalsize,labelfont=sf,textfont=sf]{subfig}
\usepackage{textcomp}
\usepackage{stfloats}
\usepackage{hyperref}
\usepackage{verbatim}
\usepackage{adjustbox}
\usepackage{booktabs}
\usepackage{cite}
\usepackage{multirow}
\usepackage{multicol}
\usepackage[para,online,flushleft]{threeparttable}
\usepackage{amsfonts}
\usepackage{pifont}
\usepackage{amsmath}
\usepackage[]{footmisc}
\usepackage{xurl}
\usepackage{xcolor}
\newcommand\hr[1]{\textcolor{red}{#1}} 
\renewcommand{\hr}[1]{#1} 

\usepackage{arydshln}
\usepackage{lineno}

\begin{document}

\begin{frontmatter}

\title{Evaluating lightweight unsupervised online IDS for masquerade attacks in CAN\tnoteref{t1}}

\tnotetext[t1]{This manuscript has been co-authored by UT-Battelle, LLC, under contract DE-AC05-00OR22725 with the US Department of Energy (DOE). The US government retains and the publisher, by accepting the article for publication, acknowledges that the US government retains a nonexclusive, paid-up, irrevocable, worldwide license to publish or reproduce the published form of this manuscript, or allow others to do so, for US government purposes. DOE will provide public access to these results of federally sponsored research in accordance with the DOE Public Access Plan (http://energy.gov/downloads/doe-public-access-plan).}

\date{}

\author[label1]{Pablo Moriano\corref{cor1}}
\ead{moriano@ornl.gov}

\author[label1]{Steven C. Hespeler}
\ead{hespelersc@ornl.gov}

\author[label2]{Mingyan Li}
\ead{lim3@ornl.gov}

\author[label3]{Robert A. Bridges}
\ead{robert.bridges@ai.se}

\cortext[cor1]{Corresponding author.}

\address[label1]{Computer Science and Mathematics Division, Oak Ridge National Laboratory, Oak Ridge, TN 37830, USA}

\address[label2]{Cyber Resilience and Intelligence Division, Oak Ridge National Laboratory, Oak Ridge, TN 37830, USA}

\address[label3]{AI Sweden, Gothenburg, Sweden}

\begin{abstract}
Vehicular controller area networks (CANs) are susceptible to masquerade attacks by malicious adversaries. In masquerade attacks, adversaries silence a targeted ID and then send malicious frames with forged content at the expected timing of benign frames. As masquerade attacks could seriously harm vehicle functionality and are the stealthiest attacks to detect in CAN, recent work has devoted attention to compare frameworks for detecting masquerade attacks in CAN. However, most existing works report offline evaluations using CAN logs already collected using simulations that do not comply with the domain's real-time constraints. Here we contribute to advance the state of the art by presenting a comparative evaluation of four different non-deep learning (DL)-based unsupervised online intrusion detection systems (IDS) for masquerade attacks in CAN. Our approach differs from existing comparative evaluations in that we analyze the effect of controlling streaming data conditions in a sliding window setting. In doing so, we use realistic masquerade attacks being replayed from the ROAD dataset. We show that although evaluated IDS are not effective at detecting every attack type, the method that relies on detecting changes in the hierarchical structure of clusters of time series produces the best results at the expense of higher computational overhead. We discuss limitations, open challenges, and how the evaluated methods can be used for practical unsupervised online CAN IDS for masquerade attacks.
\end{abstract}

\begin{keyword}
Anomaly detection \sep controller area network \sep in-vehicle security \sep machine learning \sep online algorithms \sep streaming data.
\end{keyword}

\end{frontmatter}


\section{Introduction} \label{sec:Introduction}

Today's vehicles are intricate cyber-physical systems with growing connectivity capabilities to interact with vehicle-to-everything (V2X) technologies~\citep{Kawser:2019:Perspective:V2X:5G, Minawi:2020:ML:IDS:CAN, Buscemi:2023:Survey:CAN:Reverse:Engineering}.
This increasing connectivity expands the attack surface to adversaries that can access in-vehicle networks. The Controller Area Network (CAN) is the most prevalent protocol used on in-vehicle networks allowing hundreds of electronic control units (ECUs) communicate vehicle functionality~\citep{Standard:2006:CAN:Low:Speed, Standard:2015:CAN:Data:Link:Layer, Standard:2016:CAN:High:Speed}. While the lack of security features makes CAN lightweight and inexpensive, it also makes it susceptible to attacks that aim to manipulate communications and significantly degrade in-vehicle network performance. This has been shown to result in life threatening incidents such as unintended acceleration, brake manipulation, and rogue wheel steering~\citep{Miller:2015:Remote:Exploit:Black:Hat, Miller:2016:CAN:Exploit:Details}.

CAN attacks are usually classified as fabrication, suspension, and masquerade attacks~\citep{Cho:2016:Fingerprinting:ECUs, Verma:2024:ROAD:dataset}. From these, masquerade attacks are the hardest to detect because they do not perturb the regular frame timing which can be detected using time-based methods~\citep{Woo:2014:Practical:Wireless:Masquerade:Attack, Jo:2021:CAN:Survey:And:Countermeasures}. Instead, in masquerade attacks, the adversary first silences a targeted ECU and then injects spoofed frames which alters the content of regular frames instead of time-related patterns. This makes masquerade attacks the stealthiest, often having an effect on the regular evolution of vehicle states usually measured by sensors~\citep{Akowuah:2021:Physical:Invariant:Attack:Detection}. Therefore, recent research efforts have been channeled on designing and deploying intrusion detection systems (IDS) against advanced masquerade attacks in CAN~\citep{Ansari:2017:Low:Cost:Masquerade:Attack:Detection, Moriano:2022:AHC, Shahriar:2023:CANShield, Shahriar:2023:Cantropy}.

IDS for detecting masquerade attacks in CAN are often based on unsupervised anomaly detection techniques aiming to characterize the regular relationships between physical signals in a vehicle~\citep{Moriano:2022:AHC, Sharmin:2023:Benchmark:CAN:IDS:ROAD}. Some of them focus on mining time series relationships based on deep learning (DL) models that consequently have high computational cost~\citep{Hanselmann:2020:CANet, Shahriar:2023:CANShield, Nichelini:2023:Canova:Hybrid:IDS:CAN}. In addition, most existing CAN IDS approaches are tested offline using datasets or CAN logs already collected from a real or simulated environment. This has been synthesized in existing comparative evaluation frameworks and studies in the CAN IDS space~\citep{Ji:2018:Comparative:Performance:Evaluation:IDS:CAN, Sharmin:2023:Benchmark:CAN:IDS:ROAD}. The offline evaluation approach is different from online evaluation, in that the latter is performed using streaming CAN data from a vehicle, simulation, or data log replay~\citep{Sharmin:2024:Benchmarking:CAN:IDS:Survey}. Therefore, offline evaluation hinders the understanding of realistic capabilities and limitations of CAN IDS in a streaming environment. By conducting a comparative evaluation of non DL-based unsupervised online IDS for masquerade attacks in CAN, we seek to provide insights into the effectiveness and limitations of these methods in resource-constrained and streaming environments.

In this paper, we address this gap of knowledge by introducing an empirical and unbiased comparative evaluation of four different non-DL-based unsupervised online IDS for masquerade attacks: Matrix Collelation Distribution, Matrix Correlation Correlation, Ganesan17~\citep{Ganesan:2017:Exploiting:Correlations:Heterogeneous:Sensors}, and Moriano22~\citep{Moriano:2022:AHC}. We focus on non DL-based approaches because vehicular data change over time and space. This dynamic nature makes it difficult for DL-based methods to extract relevant features, often leading to poor performance~\citep{Macas:2022:Survey:DL:Cybersecurity:Progress:Challenges:Opportunities}. Here, Matrix Correlation Distribution and Matrix Correlation Correlation methods are simple proposed baselines while Ganesan17 and Moriano22 are already proposed methods in the literature. We test them in simulated masquerade attacks from the ROAD dataset~\citep{Verma:2024:ROAD:dataset}. We rely on a sliding window approach to simulate data log replays of a continuous data stream~\citep{Babcock:2002:Sampling:From:Moving:Window, Gama:2012:Survey:Data:Streams}. \hr{Prior work has investigated how windowing strategies affect anomaly detection outcomes~\citep{Zhang:2019:PDD, Zoppi:2019:SlidingWindows}, but has not explored overlapping configurations in multivariate streaming scenarios.}

We analyze the joint effect of window size and offset on the AUC-ROC as evaluation metric for intrusion detection across the different attacks in the ROAD dataset. We also report an analysis of the testing time per window (TTW)~\citep{Nichelini:2023:Canova:Hybrid:IDS:CAN} of CAN IDS as a proxy of inference time and discuss insights into the real-time detection capability of these approaches. We further contribute to advance the state of the art in this topic by making publicly available the code of the IDS considered in this study~\citep{Moriano:2024:Online:IDS:Benchmark:Code}. This furthers research in this topic by allowing other researchers to replicate our results and compare new algorithms with the evaluated IDS.

Our results indicate that overall the evaluated unsupervised online CAN IDS are not effective in consistently detecting every attack type in the ROAD dataset. Attacks in the ROAD dataset have different levels of impact on the regular correlation patterns of signals. However, evaluation performance metrics tend to favor Moriano22 and Matrix Correlation Distribution methods. This aligns with the expectation that impacting correlations across signals tends to disrupt the hierarchy of their relationships and their pairwise similarity distribution. Notably, among the variety of attacks in the ROAD dataset, \texttt{max\_engine\_attack} and \texttt{correlated\_attack} are the ones that can be detected more easily across CAN IDS. Our results also shed light on the impact that sliding window parameters (i.e., window length and offset) have on the detection task. 



\section{Background} \label{sec:background}

\subsection{CAN Protocol} \label{subsubsec:CAN Protocol}

\begin{figure}[htb]
\centering
\includegraphics[width=1.0\columnwidth]{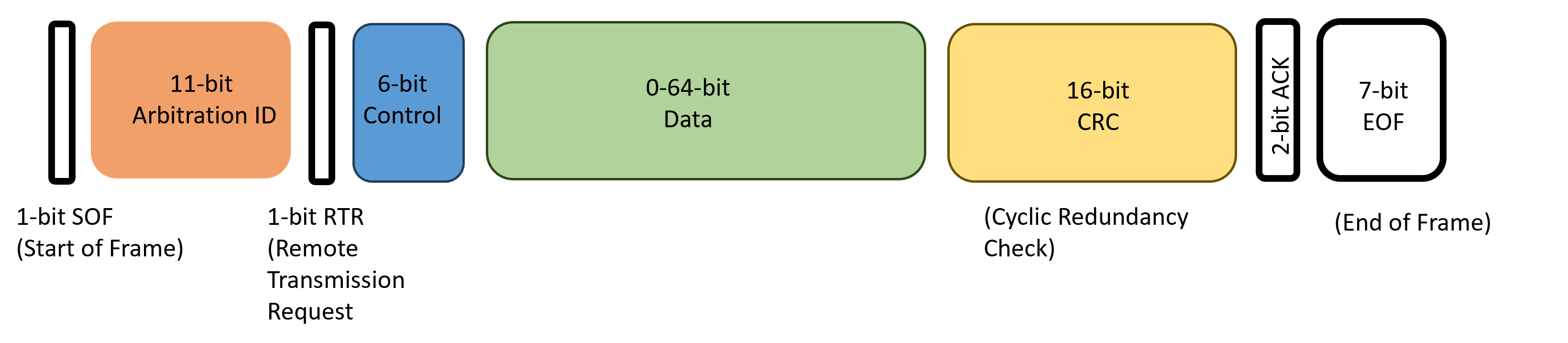}
\caption{CAN frame illustration.}
\label{fig:CAN frame}
\end{figure}

CAN is a message-based communication standard for interconnecting ECUs within industrial application and vehicles. Developed by Bosch in 1980s, CAN operates at Open Systems Connection (OSI) lowest two layers (physical and data-link layers), and is renowned for its reliability and efficiency in automotive and industrial networking applications. CAN facilitates communication among ECUs without requiring a central bus controller, using two-wire bus for seamless message transmission through frames containing arbitration, control, data, and error-checking fields, as illustrated in Fig.~\ref{fig:CAN frame}. Of particular relevance to this paper's context are two fields, the 11-bit Arbitration ID and the up-to-8-byte data field.


CAN is physically implemented as two wires. When a ECU sends a 1 bit, the voltage of the two wires are left unchanged, whereas a 0 bit is sent by manipulating the voltage of one wire up and the other down. Since this is a broadcast network (i.e., all ECUs can talk and listen on the two wires), if one ECU sends a 0, it will overwrite any 1 being sent. Hence, arbitration, a  process by which the ECUs coordinate who gets to talk is needed to avoid conflicts. The arbitration ID identifies messages on the CAN bus~\citep{Cho:2016:Fingerprinting:ECUs, Verma:2024:ROAD:dataset}. When multiple ECUs broadcast simultaneously, they send the ID value of their CAN frame one bit at a time, starting with the most significant bit. 
ECUs sending a 1 are overwritten by any ECUs sending a 0, and these overwritten ECUs cease their transmission. These ECUs then withdraw from contention and switch to receive mode. The ECU with the lowest ID wins arbitration and transmits its message. ECUs that lose arbitration retry transmission when the bus becomes available.

The data field contains the actual information being transmitted between ECUs. The payload consists of up to 8 bytes of data. Each or multiple bytes represent a specific piece of information, known as ``signals,'' such as sensor readings or status information. The data payload is essential for exchanging real-time data between ECUs. CAN frames with a certain arbitration ID are transmitted at a predetermined frequency with fixed payload structure.  For instance, the power control module transmits ID 0x102 with data field containing engine RPM, vehicle speed, and odometer signals every 0.05s \citep{Verma:2024:ROAD:dataset}.

CAN also includes a built-in sophisticated fault confinement mechanism to handle anomalies such as faulty cables, noise, or malfunctioning CAN nodes etc. Each node constantly monitors its performance and creates an ``error frame'' in response to detecting a CAN error. There are three possible CAN error states: error active state, error passive state, and error bus-off state, with the last one being prohibited from sending or receiving CAN messages. Prior work by Cho and Shin~\citep{Cho:2016:ErrorHandle} provides more details on CAN error handling.

\subsection{Threat Model} \label{subsubsec:Threat Model}
We assume a scenario where an adversary has already gained access to the CAN bus (e.g., by the OBD-II port or through wireless communication)~\citep{Checkoway:2011:Automotive:Attack:Surfaces}. Here, the adversary may compromise two ECUs, one as a \emph{strong attacker} (i.e., fully compromising an ECU including capabilities for injecting arbitrary attack messages) and the other as a \emph{weak attacker} (i.e., weakly compromising an ECU including capabilities to stop/suspend the ECU from transmitting certain messages)~\citep{Cho:2016:Fingerprinting:ECUs}. Although it is easier for an adversary to become a weak rather than a stronger attacker, researchers have demonstrated how to become strong attackers~\citep{Miller:2013:Adventures:Automotive:Networks:Control:Units, Miller:2014:Survey:Remote:Automotive:Attack:Surfaces}.

\subsection{CAN Vulnerabilities} \label{subsubsec:CAN Vulnerabilities}
Despite being robust, the CAN protocol can suffer from a multitude of cyber attacks due to its inherent vulnerabilities. 
Firstly, the absence of communication encryption protection leaves CAN messages susceptible to interception and manipulation, allowing the attackers to eavesdrop on communications or tap into and alter the network communication without being detected. 
Secondly, the lack of authentication mechanisms implies that nodes within the CAN network cannot verify the identities of message senders, making it easier for the unauthorized parties to gain access and tamper with transmission data. 
Moreover, the broadcasting nature of the CAN protocol, notably when compounded by lack of encryption and authentication, essentially allows any node to receive messages intended for others, and to inject illegitimate messages at its will. By exploiting such vulnerabilities, a malicious node can potentially manipulate message arbitration ID to disrupt message delivery priorities, launch Denial of Service (DOS), or launch impersonation attacks, to create communication errors or system malfunctions.  
Finally, CAN vulnerabilities can also lead to exploits of higher-level protocols. An example would be an attack on ECUs' Unified Diagnostic Services (UDS)\textemdash a diagnostic communication protocol used within automotive electronics, by injecting and manipulating CAN messages \citep{miller:2016:can}. 

In terms of attacks that manipulate CAN messages, there are three main types~\citep{Cho:2016:Fingerprinting:ECUs, Verma:2024:ROAD:dataset}: 
\begin{itemize}
    \item \textit{Fabrication attacks:} A strong attacker uses a compromised ECU to inject messages with malicious IDs and data fields. These IDs are often set to low values, which indicate high priority. Repeatedly injecting high-priority messages can block legitimate ECUs, causing denial-of-service (DoS) attacks or disrupting vehicle functions. 
    \item \textit{Suspension attacks:} A weak attacker compromises an ECU to stop it from sending some or all of its messages. This is done by removing ID frames, disconnecting the ECU from the CAN bus. This not only affects the compromised ECU but also disrupts receiver ECUs relying on those messages.  
    \item \textit{Masquerade attacks:} This advanced attack requires two compromised ECUs, one as a strong attacker and one as a weak attacker. \hr{The weak attacker monitors the frequency of a targeted ID and, depending on the threat model, either compromises the software of the transmitting ECU or triggers a denial-of-service condition such as a bus-off attack to suppress its transmissions. Once the legitimate messages are suppressed, a strong attacker injects messages with the same ID at the original frequency.} The ID frequency remains unchanged, but the source changes, and the messages contain forged content.
\end{itemize}
We distinguish between spoofing and masquerade attacks. In spoofing attacks, a compromised ECU sends false (spoofed) frames to trigger a reaction from legitimate ECUs, but the legitimate ECU continues to function. Existing datasets, such as HCRL Car Hacking~\citep{Song:2020:In-Vehicle:IDS:Using:CNN} and can-train-and-test~\citep{Lampe:2024:can:train:test:Curated:Dataset:Automotive:Intrusion:Detection}, include spoofing/masquerade attacks that overwhelm the CAN bus with both spoofed and legitimate messages, referred to as unsophisticated masquerade attacks~\citep{Lampe:2024:can:train:test:Curated:Dataset:Automotive:Intrusion:Detection}. These attacks make detection easier because the frequency of the targeted ID changes, but legitimate signals are not suppressed. As a result, simpler methods based on frequency detection can be used~\citep{Blevins:2021:Time:Based:CAN:IDS}. Our work focuses on sophisticated masquerade attacks from the ROAD dataset~\citep{Verma:2024:ROAD:dataset}, the only dataset, to our knowledge, that releases such attacks where legitimate signals are suppressed.



\section{Related Work} \label{sec:related-work}

Here we review and contrast previous offline comparative evaluation studies on CAN IDS (Section~\ref{subsec:Prior Work Closely Related to the Present Study}) and online CAN IDS methods (Section~\ref{subsec:Other Prior Work Related to the Present Study}).

\subsection{Prior Work Closely Related to the Present Study} \label{subsec:Prior Work Closely Related to the Present Study}

Ji et al.~\citep{Ji:2018:Comparative:Performance:Evaluation:IDS:CAN} presented a comparative analysis of four statistical-based CAN IDSes using entropy, clock skew, ID sequences, and CAN bus throughput. Compared methods relied only on ID timestamp arrivals. Their evaluation was done in simulated attack datasets. They found that the clock skew method performed best across attacks while the entropy method falls short on detecting replay attacks.

Dupont et al.~\citep{Dupont:2019:Evaluation:IDS:CAN} introduced a unifying framework for CAN IDS evaluation of time-based methods. In doing so, they used data from two live vehicles and a CAN prototype. They used both simulated attacks and publicly available datasets with attacks\citep{Lee:2017:OTIDS:Dataset}. They found that among evaluated methods, they perform well only on attacks that generate significant disruptions in CAN traffic. They suggested that methods that consume bit representations and semantics of CAN messages will perform better.

Stachowski et al.~\citep{Stachowski:2019:Assessment:CAN:IDS:Performance} introduced a framework for evaluating CAN IDS. Their approach focused on the online evaluation of three existing solutions from identified vendors, so that CAN IDS are tested in real vehicles for real-time intrusion detection. Targeted ID attacks were introduced in test vehicles while vehicles were in motion and stationary. None of the tested CAN IDS was found effective for detecting the variety of attacks.

Blevins et al.~\citep{Blevins:2021:Time:Based:CAN:IDS} focused on evaluating time-based CAN IDS on the fuzzy and targetted ID attacks from the ROAD dataset. Compared methods include mean inter-message time, fitting a Gaussian curve, kernel density estimation, and binning. They found that the binning method outperformed the other methods, specially those fitting a distribution on inter-arrival times, in terms on AUC-ROC, AUC-PR, and F1 score. They reported a latency analysis for the binning algorithm.

Agbaje et al.~\citep{Agbaje:2022:Framework:CAN:IDS:Evaluation} introduced a framework for comprehensive performance evaluation analysis of eight state-of-the-art CAN IDS. The proposed framework provides a consistent methodology to evaluate and asses CAN IDS. They focus on using previously published datasets recorded from a real vehicle and report traditional classifier performance evaluation metrics to reduce uncertainty when comparing IDS approaches from different sources.

Pollicino et al.~\citep{Pollicino:2023:Performance:Timing:IDS:Comparison} introduced an experimental comparison on the performance evaluation of eight CAN IDS algorithms over two different datasets, an in-house one and one that is already publicly available (i.e., OTIDS~\citep{Lee:2017:OTIDS:Dataset}). The algorithms they evaluated consume only time-based information from CAN frames as they are intended for detection of fabrication attacks. They open-sourced code implementations of these algorithms to allow reproducibility of research results and detail an unbiased experimental comparison.

Sharmin et al.~\citep{Sharmin:2023:Benchmark:CAN:IDS:ROAD} reported results of a comparative evaluation of four statistical and two ML-based CAN IDS in the ROAD dataset, including masquerade attacks. In addition to reporting traditional evaluation metrics such as accuracy, precision, recall, and F1 score, they included balanced accuracy, informedness, markedness, and Matthews correlation coefficient (MCC) which are better suited for handling imbalanced datasets~\citep{Chicco:2020:MCC:Over:F1}. They also reported training and testing time for each IDS. They found that frequency-based approaches generally perform well at detecting fabrication attacks while other algorithm categories did not perform well based on low MCC scores.

Compared to the comparative evaluation studies mentioned above, the present paper is unique because it focuses on online IDS methods for masquerade attacks. Note that previous comparative evaluation studies focused on: (1) comparing CAN IDS methods in an offline fashion (see works~\citep{Ji:2018:Comparative:Performance:Evaluation:IDS:CAN, Dupont:2019:Evaluation:IDS:CAN, Blevins:2021:Time:Based:CAN:IDS, Pollicino:2023:Performance:Timing:IDS:Comparison, Sharmin:2023:Benchmark:CAN:IDS:ROAD}); and (2) performing evaluation tests only on fabrication attacks (see works~\citep{Ji:2018:Comparative:Performance:Evaluation:IDS:CAN, Dupont:2019:Evaluation:IDS:CAN, Stachowski:2019:Assessment:CAN:IDS:Performance, Blevins:2021:Time:Based:CAN:IDS, Agbaje:2022:Framework:CAN:IDS:Evaluation, Pollicino:2023:Performance:Timing:IDS:Comparison}). By evaluating CAN IDS methods on these two key aspects, we contribute to fill this gap of knowledge by analyzing the performance of unsupervised online CAN IDS in online fashion and testing them on the most stealthy CAN bus attacks, i.e., masquerade attacks. Our contribution enriches the state of the art by introducing a more realistic assessment of CAN IDS behavior in a realistic but constrained environment.

\subsection{Other Prior Work Related to the Present Study} \label{subsec:Other Prior Work Related to the Present Study}

Desta el al.~\citep{Desta:2020:ID:Sequence:Analysis:Intrusion:Detection} proposed a CAN IDS based on a LSTM sequence predictor model focused on predicting ID sequences. They performed in-house collection of CAN data using a real car. They replayed a CAN data log using the socketCAN API to evaluate their CAN IDS. They found that they can obtain better results by using the log loss of the predicted ID and the true ID.

Sunny et al.~\citep{Sunny:2020:Hybrid:Approach:AnomalyDetection:CAN} proposed a hybrid CAN IDS that leverages patterns of recurring messages and time interval between them. The proposed approach produces fast inference times being informed by best and worse time intervals between two consecutive frames. As CAN messages are expected to be published every 2 ms, they conclude that the proposed method can be used for real-time settings on simulated attack scenarios.

Jedh et al.~\citep{Jedh:2022:Evaluation:Architectures:IDS:CAN} evaluated four architecture designs for real-time CAN IDS under malicious speed reading message injections. They focus on a single ML-based IDS that captures the pattern of the sequences of CAN messages represented with a direct graph. They showed that the optimal architecture for deploying a real-time CAN IDS is based on two processes, a process for CAN monitoring and another one exclusively for anomaly detection. 

Jadidbonab et al.~\citep{Jadidbonab:2022:Realtime:Testbed:ML:Trainig:Validation} introduced an automotive security testbed that combines a full-scale in-vehicle network (i.e., vector CANoe) and a car simulator (i.e., CARLA) to generate network traffic in realistic scenarios. They tested ML-based IDS on their testbed and found that online algorithms performed poorer when compared to training offline classifiers on previously collected logs.

Marfo et al.~\citep{Marfo:2024:Detecting:Masquerade:Attacks:CAN} introduced a graph ML framework for detecting masquerade attacks in the CAN bus. They combined shallow graph embeddings with time series features to enhance the detection of masquerade attacks. They show that this approach leads to statistically significant improvements in the detection rates of masquerade attacks compared to a baseline that uses only graph-based features.

Recent advances in unsupervised anomaly detection for tabular~\citep{Zhao:2019:PyOD, Zoppi:2025:ConfidenceEnsembles} and image data~\citep{Guo:2025:Dinomaly} offer promising directions, but are not directly applicable to the multivariate streaming time series setting considered here.

\section{Methods} \label{sec:Methods}

Common to all the methods evaluated in this work is our focus on processing a set of time series representing signals in the CAN bus. For that, we assume that we have access to the time series representation obtained from CAN logs captured during a vehicle's drive. Recall that the subsequent methods are expected to work in online fashion. This means that they 
process a time series stream to  provide near real-time alerts. 

Although the terms streaming and online algorithms are sometimes used interchangeably in the literature, we draw a distinction in line with prior surveys on data stream mining and adaptive learning~\citep{Gama:2012:Survey:Data:Streams, Gama:2014:Survey:Concept:Drift, Avogadro:2020:Online:Anomalies:Time:Series}. In our context, streaming algorithms refer to those that process each data instance individually as it arrives, typically under strict memory and time constraints. In contrast, we use online algorithms to describe methods that process small batches or windows of data incrementally, allowing for limited buffering and adaptation over time. Throughout this work, we adopt the latter interpretation. When we refer to online algorithms, we mean algorithms designed to operate on continuously arriving data by processing small, sequential segments of the stream, updating models and decisions incrementally. Note that our inclination to comparative evaluation unsupervised learning methods is guided in part by the fact that it is challenging to apply supervised methods on streaming data due to their static nature~\citep{Ferragut:2012:New:Principled:Approach:Anomaly:Detection}. 

\hr{Therefore, in this work, we follow Gama et al.~\citep{Gama:2012:Survey:Data:Streams, Gama:2014:Survey:Concept:Drift} and Avogadro et al.~\citep{Avogadro:2020:Online:Anomalies:Time:Series} in defining online anomaly detection as methods that operate on sequentially arriving data, processing it in small windows without requiring access to future observations. Our framework adheres to this setup and does not assume knowledge of future windows at inference time.}



\subsection{Time Windows} \label{subsubsec:Time Windows}

\begin{figure}[htb]
\centering
\includegraphics[width=1.0\columnwidth]{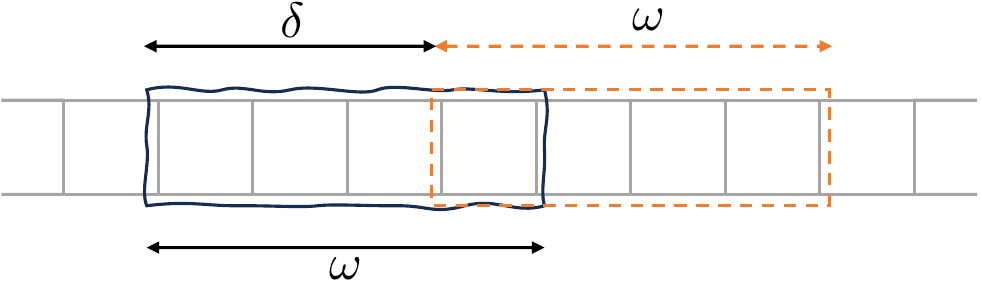}
\caption{Sliding windows concept. The stream of time series data is partitioned using windows of length $\omega$. The sliding windows step, or offset, is $\delta$. }
\label{fig:sliding window}
\end{figure}

The presented comparative evaluation is based on sliding time windows
of the stream of time series data. 
The assumption behind using windows is that more recent information is more relevant for decision-making than past data. In particular, we focus on using \emph{sliding windows} of fixed size for anomaly detection~\citep{Qin:2019:Symmetry:Degree:Measurement:Anomaly:Detection}. The windows are defined in terms of the number of observations or sequence-based windows~\citep{Babcock:2002:Sampling:From:Moving:Window}. This involves a sequence of $\omega$-length windows (i.e., $\omega \in \mathbb{Z}$) $W=\{W_{j} : j \ge 1\}$ with a sliding step (or offset $\delta \in [1, \omega]$) and observations arriving sequentially. Let $Y_{t}$ for $t \ge 0$ be a set of time series in a given window $W$, then $Y_{t}$ contains time series over the interval $[\delta \times (j - 1), \delta \times (j - 1) + \omega]$. Hence, $Y_{t}$ is a set of time series of the most recent $\omega$ observations up to time $t$ with $\delta$ observations that expire. Figure~\ref{fig:sliding window} shows this concept visually.

\subsection{Experimental Design} \label{subsubsec:Experimental Design}

\begin{figure*}[!hbp]
\centering
\includegraphics[width=1.0\textwidth]{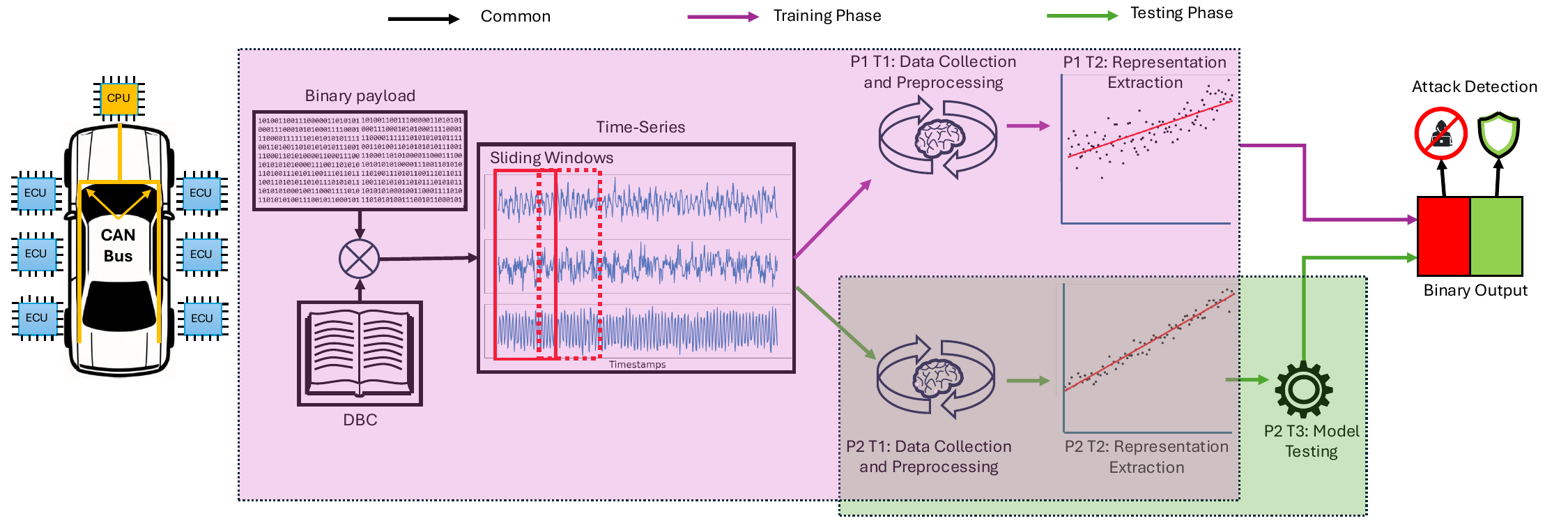}
\caption{An overview of our proposed comparative evaluation framework.}
\label{fig:Design Diagram}
\end{figure*}

Here we describe our proposed comparative evaluation framework. More details can be found in Fig.~\ref{fig:Design Diagram}. Our proposed framework has two phases: i) $P_{1}$: training phase and ii) $P_{2}$: testing phase. These phases have some sequential tasks. Some of them are unique to one phase while others are common to both phases. The tasks are: i) $T_{1}$: data collection and preprocessing, $T_{2}$: \hr{representation extraction}, and $T_{3}$: model testing. $P_{1}$ consists of tasks $T_{1}$ and $T_{2}$, whereas $P_{2}$ consists of tasks $T_{1}$, $T_{2}$, and $T_{3}$. 

The proposed comparative evaluation framework uses time windows for data processing and attack detection within the same time window. This means that our comparative evaluation is focused on detecting attacks at the window level rather than at the message level. \hr{Note that both phases can executed on edge-capable devices such as the Raspberry Pi or Nvidia Jetson Nano. While these are not automotive-grade embedded systems, they are widely used proxies in IDS research for evaluating feasibility under constrained compute and memory budgets~\citep{Shahriar:2023:CANShield}. The average processing times reported in Table~\ref{table: TTW summary table} provide a proxy for lightweight deployability.} We now explain the technical aspects of the framework in more detail. 

\subsubsection{Training Phase (P1)} \label{subsubsec:Training Phase}

\emph{1) Data Collection and Preprocessing (T1):} \label{subsubsec:Data Collection and Preprocessing P1}

The first task common to all the algorithms in our comparative evaluation is the configuration of the data collection and preprocessing pipeline. We assume that the evaluated algorithms run on light-weight devices connected to the CAN bus through the OBD-II port or a dedicated ECU with direct access to the CAN bus. This setup allows continuous collection of raw CAN messages along with its translation from binary to time series using its corresponding decoding instructions. Note that there are available open-source CAN data loggers such as SavvyCAN and SocketCAN and commercial tools such as CANalyzer and VehicleSpy. We also assume that for translating CAN binary payloads to multidimensional time series, algorithms are loaded with OEM's DBC file or use CAN decoding software such as LibreCAN, CAN-D, or CANMatch~\citep{Pese:2019:LibreCAN, Verma:2021:CAN-D, Buscemi:2021:CANMatch}. For a more comprehensive discussion on CAN signals reverse engineering, we refer the reader to the survey by Buscemi et al.~\citep{Buscemi:2023:Survey:CAN:Reverse:Engineering}. 

We focus on processing a set of $n$ signals, i.e., $\mathcal{S} = \{s_{1}, s_{2}, \ldots, s_{n}\}$. Here $n$ means the total number of decoded signals. As each of these time series is generated at a characteristic frequency (imposed by their corresponding ID), we linearly interpolated them in common timesteps to have the same frequency producing evenly spaced time series. We use a common frequency of 100Hz to take into account the fact that most IDs do not send messages above this frequency threshold. This ensures that $\forall s_{i} \in \mathcal{S}, |s_{i}| = t_{0}$, where $t_{0}$ is the total number of time steps in the training set.

This multivariate time series data is stored in a matrix $\mathbf{X}_{train} \in \mathbb{R}^{t_{0} \times n}$. On this matrix, we discard any constant time series and apply min-max normalization to the remaining time series. 

\emph{2) \hr{Representation Extraction} (T2):} \label{subsubsec:Model Fitting P1}

We fit data-driven models for each of the evaluated methods using the training data, i.e., $\mathbf{X}_{train}$. Fitted models in this task are then used as the regular (normal) state representation from which we derive comparisons with models fitted at the testing stage. The details of the fitted models for each of the becnhmarked methods are in Section~\ref{subsec:Evaluated Algorithms}.

\subsubsection{Testing Phase (P2)} \label{subsubsec:Testing Phase}


The testing phase executes similar tasks as already discussed in the training phase (see tasks T1 and T2 in Section~\ref{subsubsec:Training Phase}) and adds T3. Note that while P1 is run only once, the tasks in P2 are run continuously and sequentially as new messages are processed and there is a need to check for anomalies. We detail here changes with respect to specific parameters.

\emph{1) Data Collection and Preprocessing (T1):} \label{subsubsec:Data Collection and Preprocessing P2}

In the testing stage, data is processed using a sliding window approach (see more details in Section~\ref{subsubsec:Time Windows}). Specifically, smaller sections of $\mathbf{X}_{test}$ of length $\omega$ are selected to create a view of the steaming ($\in \mathbb{R}^{\omega \times n}$). 
Evaluated methods consume this batch of data to arrive at conclusions. 

\emph{2) \hr{Representation Extraction}  (T2):} \label{subsubsec:Model Fitting P2}

\hr{During the testing phase (P2), we apply the evaluated methods to extract statistical representations from the windowed test data $\mathbf{X}_{test}$ and compare them to those obtained from the training data (P1). No model fitting is performed on the test set. The evaluated methods are discussed in Section~\ref{subsec:Evaluated Algorithms}.} 

\emph{3) Model Testing (T3):} \label{subsubsec:Model Testing P2}

This step entails a data-driven decision-making for deciding if a particular window is containing an anomaly or not. The specifics of the criteria for detection are discussed in Section~\ref{subsec:Evaluated Algorithms}. 

\subsection{Evaluated Algorithms} \label{subsec:Evaluated Algorithms}

This subsection describes in detail each of the four algorithms used in this research. Note that we focus on non DL-based algorithms because the dynamic nature of CAN data makes it hard for DL-based methods to extract useful features, which reduces their performance~\citep{Macas:2022:Survey:DL:Cybersecurity:Progress:Challenges:Opportunities}. For the sake of readability, we associate each algorithm name of the first author and the last two digits of the publication year (when applicable). Specifically, we describe the following methods: Matrix Correlation Distribution (see Section~\ref{subsubsec:Matrix Correlation Distribution}), Matrix Correlation Correlation (see Section~\ref{subsubsec:Matrix Correlation Correlation}), Ganesan17~\citep{Ganesan:2017:Exploiting:Correlations:Heterogeneous:Sensors} (see Section~\ref{subsubsec:Genesan Method}), and Moriano22~\citep{Moriano:2022:AHC} (see Section~\ref{subsubsec:Moriano Method}). 

We notice that none of the evaluated algorithms has been distributed with a corresponding implementation. In addition, many of their implementation details are not provided. To address these limitations, we make available our referenced implementations and provide a detailed description of the assumptions and parameters tested in each of the algorithms. This allows practicioners in the area the opportunity to easily replicate our experiments and comparative evaluation novel signal-based detection algorithms with respect to the state of the art. Note that becnhmarked algorithms produce anomaly scores in the range 0 to 1 that is being used for computing evaluation metrics.

All method computes pairwise Pearson correlations among time series in $\mathbf{X}_{train}$ (and $\mathbf{X}_{test}$) obtaining $\mathbf{R}_{train} \in \mathbb{R}^{n \times n}$ (and corresponding $\mathbf{R}_{test} \in \mathbb{R}^{n \times n}$), symmetric matrices whose $r_{ij}$ entry is the Pearson correlation coefficient~\citep{Pearson:1895:Pearson:Correlation} between the signals $s_{i}$ and $s_{j}$. 
Pearson correlation values that are close to $\pm 1.0$ indicate strong positive or negative correlation. Note that time series that have positive correlations are expected to move jointly (i.e., when one time series increases or decreases, the other time series also increases or decreases). As vehicle dynamics are constrained by laws of nature, we expect that: (1) clusters of correlated signals emerge, for example, as there is an increase in the speed of the vehicle, there is a subsequent increase in the speedometer reading and the speed of the four wheels, and (2) such clusters are disturbed upon a cyberattack given that time series correlations could be broken or changed significantly. 

We let $\mathbf{U}$ denote the strictly upper triangular matrix of $\mathbf{R}$, i.e., $u_{ij} = r_{ij}$ if $i<j$ and $u_{ij} = 0$ if $i\geq j$. 
We create strictly upper triangular matrices $\mathbf{U}_{train}, \mathbf{U}_{test}$, respectively, for  both training and testing correlation matrices ($\mathbf{R}_{train}$ and $\mathbf{R}_{test}$) to estimate pairwise similarity distributions among time series. 

Note that the evaluated algorithms suit streaming environments. They process small data batches and do not rely on large, representative training datasets, unlike DL-based methods.

\subsubsection{Implementation of Matrix Correlation Distribution} \label{subsubsec:Matrix Correlation Distribution}
This method seeks statistical significance of the signals' pairwise Pearson correlations. 
Using $\mathbf{U}_{train}, \mathbf{U}_{test}$ as above, we use the Mann-Whitney U test~\citep{Mann:1947:MannWhitneyUtest} $p$-value  
to create an anomaly probability score; specifically, one minus the $p$-value is the anomaly score so that a larger score is more 
anomalous. 
The Mann-Whitney U test is a non-parametric test often used to test the difference in location between distributions. 
As the Mann-Whitney U test null hypothesis is that the Pearson similarity distribution underlying benign conditions and attack conditions is the same, we use one minus the $p$-value as the anomaly score. 
Intuitively, higher $p$-values suggest not rejecting the null hypothesis; this translates to a lower anomaly probability score. Conversely, lower $p$-values suggest rejecting the null hypothesis; this translates to a higher anomaly probability score. 

\subsubsection{Implementation of Matrix Correlation Correlation} \label{subsubsec:Matrix Correlation Correlation}
This method considers the upper triangular entries of $\mathbf{U}_{train}, \mathbf{U}_{test}$ (the signals' pairwise Pearson correlation coefficients) and computes the  statistical significance of their \textit{Spearman correlation} \citep{Spearman:1987:Spearman:Correlation}, a measure in $[-1,1]$ of their members ordinal rank.   
Simply unravel the upper triangular entries of $\mathbf{U}_{train}, \mathbf{U}_{test}$ into two sequences and compute their  Spearman correlation.
Note that Spearman correlation furnishes, alongside the correlation value, a $p$-value giving the likelihood that the two input vectors are from uncorrelated systems (so low $p$-value corresponds to the two input vectors coming from correlated systems).  
We focus on the $p$-value of the Spearman correlation as this captures the reliability of the correlation. The Spearman correlation coefficient is better suited for non-normally distributed continuous data, ordinal data, or data with potential outliers as a measure of monotonic association~\citep{Schober:2018:Correlation:Interpretation}. Pearson correlation is also scaled in the range $\pm 1$ indicating a constantly increasing or decreasing monotonic association while $0$ indicating no monotonic association. 
The Spearman correlation null hypothesis is that there is no correlation between the strictly upper triangular correlations in training and testing (opposite to the first test above); hence, we use the $p$-value as the probability score. 
Here, higher $p$-values suggest not rejecting the null hypothesis; this translates to a higher anomaly score. Conversely, lower $p$-values suggest rejecting the null hypothesis; this translates to a lower anomaly score.

\subsubsection{Implementation of Ganesan17} \label{subsubsec:Genesan Method}
This method focuses on computing clusters of time series defining a context (such as aggressive or sudden driving). Note that this method does not rely on cluster characterization during training and only relies on processing testing windows. Specifically, for each time window of length $\omega$, the DBSCAN clustering method~\citep{Ester:1996:DBSCAN} is applied to a distance matrix $\mathbf{D}_{test} \in \mathbb{R}^{n \times n}$ computed as $2 * (1 - \mathbf{R}_{test})$, i.e., a squared Euclidean distance to identify clusters~\citep{CV:2012:Correlation:To:Distance}. DBSCAN is a density-based non-parametric clustering algorithm that works by grouping points with many nearby neighbors and marking as outliers the points that lie alone in low-density regions. 
By analyzing clusters of time series as opposed to individual pairs, this method aims to capture richer relationships between 
signals. 

For detection, once clusters of time series have been identified, pairwise cross correlations among time series within clusters are computed and compared with those expected for that cluster. 
Specifically, for each correlation pair, the deviation from the mean correlation value for that cluster is computed. Among each possible pairs, the maximum deviation (or error) is extracted and compared in terms of the number of standard deviations from the mean. Among all the errors in a single cluster, only the maximum error among clusters is used as reference. Finally, among all the errors among clusters, the probability of being anomalous is estimated based on the normal distribution approximation of maximum errors evaluated at a particular error, i.e., the probability that the distribution of maximum errors will take a value less or equal than the error.

\subsubsection{Implementation of Moriano22} \label{subsubsec:Moriano Method}

This method focuses on computing the similarity of hierarchical clusterings of time series in training and testing. 
Specifically, pairwise correlations matrices $\mathbf{R}_{train}$ and $\mathbf{R}_{test}$ are used to feed an agglomerative hierarchical clustering (AHC) algorithm ~\citep{Jain:1999:Data:Clustering:Survey}. We used a Ward's linkage as a metric for dissimilarity between clusters as it maximizes detection results as in the original paper~\citep{Moriano:2022:AHC}. The output of the ACH is a hierarchical clustering or a hierarchy of clusters (i.e., a set of nested clusters that are organized in a tree-like diagram known as dendrogram).  

For detection, in each time window of length $\omega$, we compute the clustering similarity $s$ between the hierarchical clustering from training (computed once) and the hierarchical clustering computed during that specific time window. Here, similarity is used as a proxy of the distance between both hierarchical clusterings. 
We compute the similarity between hierarchical clustering using the CluSim method~\citep{Gates:2019:Clusim:Package} based on the work by Gates et al.~\citep{Gates:2019:ECS}. 
The similarity value provided by this method, $s$, lies in the range $[0,1]$, where 0 implies maximally dissimilar clusters, and 1 corresponds to identical clusters. Therefore, we report the anomaly score (or probability of detection) as $1 - s$. 

\subsubsection{Algorithms' Hyperparameter Setup} \label{subsubsec:Algorithms' Hyperparameter Setup}
Matrix Correlation Distribution and Matrix Correlation Correlation algorithms do not require hyperparameters. They report anomaly scores based on $p$-values from the Mann-Whitney U test and Spearman correlation, respectively. For Ganesan17, we used the similar values as in Ganesan et al.~\citep{Ganesan:2017:Exploiting:Correlations:Heterogeneous:Sensors}, i.e., we set both $eps$ and $min\_samples$ to 1 to ensure a reasonable similarity between time series to be consider as part of the same cluster and allow individual time series to be part of its own cluster, respectively. For Moriano22, we used the sames values as in Moriano et al.~\citep{Moriano:2022:AHC}, i.e., setting $r=-5.0$ and $\alpha = 0.9$. Here, $r$ is a scaling parameter that defines the relative importance of memberships at different levels of the hierarchy. That is, the larger $r$, the more emphasis on comparing lower levels of the dendrogram (zoom in). In addition, $\alpha$ is a parameter that controls the influence of hierarchical clusterings with shared lineages. That is, the larger $\alpha$, the further the process will explore from the focus data element, so more of the cluster structure is taken into account into the comparison.

\subsection{Dataset} \label{subsubsec:Datasets}
We ran the proposed comparative evaluation on the ROAD dataset~\citep{Verma:2024:ROAD:dataset}. The ROAD dataset is a publicly available dataset for CAN IDS that contains benign and attack CAN data collected from a real vehicle. An important feature of ROAD is that it provides samples of masquerade attacks which are the types of attacks we are interested in detecting with the evaluated methods. Note that masquerade attacks in ROAD are identical to the corresponding targeted ID captures but with the ambient frames of the targeted ID removed during the attack period to simulate the masquerade attack. The ROAD dataset also provides already translated CAN time series from log files. By using ROAD in this study, we ensure comparing the performance of proposed methods with realistic, verified, and labelled attacks as opposed to synthetic ones. This allows performing the comparison under realistic conditions. 

We used the longest benign capture in the ROAD dataset (i.e., \texttt{ambient-highway-street-driving-long}) for learning the normal state of the system ($\approx$ 1 hour of data). We report results on the proposed comparative evaluation in each of the masquerade attacks provided in the ROAD dataset. Note that each of the masquerade attack files in ROAD contains hundreds of time series coming from hundreds IDs each of them with up to a few dozen signals. Table~\ref{Table: data description} shows the specific files we used from ROAD. In particular, we used the following attacks in increasing level of detection difficulty: correlated signal, max speedometer, max engine coolant temperature, reverse light on, and reverse light off. In the correlated signal attack, the correlation of the four wheel speed values is manipulated by changing their individual values. In the max speedometer and max engine coolant attacks, the speedometer and coolant temperature values are changed to their maximum. In the reverse light attacks, the state of the reverse light is modified so that it does not match the current gear of the vehicle (i.e., reverse light is off when the vehicle is in reverse and the reverse light is on when the vehicle is in forward gear).
{\color{black}\begin{table}[h]
\color{black}\centering
\caption{\color{black}Masquerade attack characteristics in the ROAD dataset~\citep{Verma:2024:ROAD:dataset}.}
\label{Table: data description}
\begin{adjustbox}{max width=0.5\textwidth}
\begin{tabular}{@{}lccp{4.1cm}@{}}
\toprule
\textbf{Attack Name} & \textbf{Duration (s)} & \textbf{Injection Interval (s)} & \textbf{Description} \\
\midrule

\texttt{correlated\_signal\_1} & 33.10 & [9.19, 30.05] & \multirow{3}{4.1cm}{Injects varying values for wheel speeds, causing the vehicle to halt.} \\
\texttt{correlated\_signal\_2} & 28.23 & [6.83, 28.23] & \\
\texttt{correlated\_signal\_3} & 16.96 & [4.32, 16.96] & \\
\hdashline

\texttt{max\_engine\_coolant} & 25.88 & [19.98, 24.17] & Injects the maximum value, triggering the coolant warning light. \\
\hdashline

\texttt{max\_speedometer\_1} & 88.02 & [42.01, 66.45] & \multirow{3}{4.1cm}{Injects the maximum value to be displayed on the speedometer.} \\
\texttt{max\_speedometer\_2} & 59.70 & [16.01, 47.41] & \\
\texttt{max\_speedometer\_3} & 86.77 & [9.52, 70.59] & \\
\hdashline

\texttt{reverse\_light\_off\_1} & 28.11 & [16.63, 23.35] & \multirow{3}{4.1cm}{Toggles the reverse light irrespective of the actual gear position.} \\
\texttt{reverse\_light\_off\_2} & 40.67 & [13.17, 36.88] & \\
\texttt{reverse\_light\_off\_3} & 57.88 & [16.52, 40.86] & \\
\hdashline

\texttt{reverse\_light\_on\_1} & 54.85 & [18.93, 38.84] & \multirow{3}{4.1cm}{Toggles the reverse light irrespective of the actual gear position.} \\
\texttt{reverse\_light\_on\_2} & 72.02 & [20.41, 57.30] & \\
\texttt{reverse\_light\_on\_3} & 64.26 & [23.07, 46.58] & \\

\bottomrule
\end{tabular}
\end{adjustbox}
\end{table}

\subsection{Evaluation Metrics} \label{subsubsec:Evaluation Metrics}

This study focuses on comparing the detection capability of the evaluated algorithms on the ROAD dataset. As such, we report summarized results over a range of detection thresholds and focus on the area under the receiver operating characteristic curve (AUC-ROC) metric. The AUC-ROC reflects the overall performance of the evaluated algorithms based on the variation of the true positive rate (TPR) with respect to the false positive rate (FPR) at various thresholds~\citep{Fawcett:2006:Introduction:ROC}. Mathematically, it is expressed as
\begin{equation*} \label{eq:AUC-ROC}
\text{AUC-ROC} = \int_{0}^{1} \text{TPR}(\text{FPR}^{-1}(x)) dx.
\end{equation*}

AUC-ROC takes values in the interval $[0, 1]$. In particular, an algorithm whose predictions are $100\%$ wrong has an AUC-ROC of $0.00$; while an algorithm whose predictions are $100\%$ correct has AUC-ROC of $1.00$. A reasonable way of interpreting AUC-ROC is as the probability that the detection algorithm ranks a random positive window more highly than a random negative window, which is desirable in our case. Note that although the AUC-ROC provides an overly estimated view in highly imbalance datasets~\citep{Davis:2006:ROC:PRC:Relationship}, we stick to AUC-ROC as the proportion of time windows overlapping with attack intervals is not imbalanced (i.e., for most of the window length and offset combinations is above $40\%$).

\hr{While alternative classification metrics (e.g., F1-score, MCC) are valuable in fixed-threshold offline evaluations, we intentionally focus on metrics best aligned with streaming contexts. Our public codebase allows easy computation of these additional metrics using preferred thresholding strategies.}




\subsection{Performance Metrics} \label{subsubsec:Performance Metrics}

For assessing the real-time capabilities of the evaluated algorithms, we also report the latency that these methods have to infer detection decisions on the sliding windows being processed. In doing so, we report the testing time per window (TTW), for each combination of window length ($\omega$) and offset ($\delta$), following the concept of testing time per packet used by Nichelini et al.~\citep{Nichelini:2023:Canova:Hybrid:IDS:CAN}. We define TTW as 
\begin{equation*} \label{eq:TTP}
\text{TTW} = \frac{\text{Total Detection Time}}{\text{Number of Windows}}
\end{equation*}
Our experiments were conducted on a AMD EPYC 7702 64-Core Processor Central Processing Unit (CPU), 256 GB of RAM, running Ubuntu 22.04.4 LTS (Jammy Jellyfish). TTW provides a way to estimate the applicability of the evaluated algorithms in a real-time setting.

\section{Results} \label{sec:Results}

In this section, we detail the experimental evaluation of the evaluated algorithms on the ROAD dataset. All methods were implemented in \texttt{Python 3.8.12}. We use the \texttt{sklearn.metrics} module to compute the evaluation metrics and the \texttt{timeit} module to compute the TTW of each evaluated algorithm. To enable the reproducibility of the evaluated methods, we make the code available in GitHub.\footnote{\url{https://github.com/pmoriano/benchmarking-unsupervised-online-IDS-masquerade-attacks}} We conduct three different but complementary analysis.

First, we compute and contrast the detection performance (based on AUC-ROC) of each algorithm per attack category in the ROAD dataset. In doing so, we display heatmaps (Section~\ref{subsec:Evaluation:Metrics:Contrast}). We hypothesize that each algorithm is better suited for detecting specific attacks in the ROAD dataset and that there is a range of $(\omega, \delta)$ combinations that maximizes AUC-ROC.

Second, we summarize evaluation metrics using summary statistics derived from the heatmaps (Section~\ref{subsec:Evaluation:Metrics:Summary}). Specifically, we report the mean ($\mu$), standard deviation ($\sigma$), median ($\eta$), minimum ($\min$), including the combination of window length and offset that produces it, and maximum ($\max$), including the combination of window length and offset that produces it. This analysis supports our hypothesis that there are better suited detection algorithms for each attack category. Overall, Moriano22 detection algorithm produces, on average, the best detection performance among attacks in the ROAD dataset. In contrast, Ganesan17 detection algorithm produces the worst detection performance among attacks in the ROAD dataset. Furthermore, we found that the easiest attacks to detect are the \texttt{max\_engine\_coolant\_attack} and \texttt{correlated\_attack}. Furthermore, we find that the \texttt{max\_speedometer\_attack} and \texttt{light\_on\_attack} are the hardest to detect. We also optimize the hyperparameters for the best-performing method, i.e., Moriano22. Using grid search, we demonstrate the potential for AUC-ROC improvement through hyperparameter tuning. 

Lastly, we summarize TTW as a proxy of performance (Section~\ref{subsec:Perfromance:Metrics:Summary}). In particular, we report the mean ($\mu$), standard deviation ($\sigma$), median ($\eta$), minimum ($\min$), including the combination of window length and offset that produces it, and maximum ($\max$), including the combination of window length and offset that produces it. This analysis reveals that more sophisticated algorithms (e.g., Ganesan17 and Moriano22) incur performance overhead for producing better detection evaluation metrics. Overall, Matrix Correlation Distribution produces, on overage, the quickest TTW among attacks in the ROAD dataset. In contrast, the Moriano22 detection algorithm runs at least four times slower, as it uses a standard AHC implementation rather than a faster approximate alternative~\citep{Sieranoja:2025:Fast:Agglomerative:Clustering:Approximation:TSP}. It is worth noting that detection algorithms that run faster (i.e., Matrix Correlation Distribution and Matrix Correlation Correlation) perform even better than Ganesan17 (see Table~\ref{table: TTW summary table}). We did not found significant differences on TTW per attack categories.

\subsection{Evaluation Metrics Contrast} \label{subsec:Evaluation:Metrics:Contrast}

\hr{To evaluate how detection metrics are influenced by sliding window configurations, we conducted a fine-grained parameter sweep over key window parameters (e.g., window size $\omega$ and offset $\delta$). These detailed visualizations are important for diagnosing parameter sensitivity and robustness, which are critical in online streaming contexts.}

Figs.~\ref{fig:distribution method}, \ref{fig:correlation method}, \ref{fig:dbscan method}, \ref{fig:ahc method} show the detection performance of each detection algorithm (i.e., Matrix Correlation Distribution, Matrix Correlation Correlation, Ganesan17, and Moriano22, respectively). Each cell in the heatmaps depicts a single AUC-ROC value for a specific combination of $\omega$ and $\delta$. Note that attack types are displayed horizontally. As attack types have different number of associated captures (see Table~\ref{Table: data description}), we compute the average in each attack category and report a single heatmap summarizing the performance in a specific category. As the standard deviation from this aggregation is negligible with respect to the average, we do not present associated figures here. The results are rounded at two decimal places. Specifically, we vary the window length (i.e., $\omega$ from 50 to 400 samples in increments of 50 samples) and the offset (i.e., $\delta$ from 10 to $\omega$ in increments of 10 samples).

Masquerade attacks in the ROAD dataset are sustained and stealthy, lasting about 17–72 seconds (see Table~\ref{Table: data description}). We tuned the sliding window parameters ($\omega$ and $\delta$) to match attack duration while keeping computation manageable. For example, a window size of 400 samples gives enough temporal context to detect slow-developing anomalies. An offset of 380 samples allows overlap between windows, which helps avoid missed detections at window edges. This offset does not delay response but balances detection coverage and computational cost. Smaller offsets (e.g., 100) increase load, while larger ones (e.g., 400) may miss attacks. These parameters match the timing of real attacks. Note that masquerade attacks do not show up in single CAN frames. They imitate normal traffic in short bursts but become apparent as gradual deviations. As shown by Jedh et al.~\citep{Jedh:2021:Detection:Message:Injection:Attacks:CAN:Using:Similarities:MSGs}, detecting these attacks requires looking across multiple frames, not just isolated ones.

Fig.~\ref{fig:distribution method} shows results for the Matrix Correlation Distribution Method. We found that overall, this method tends to perform better at detecting the \texttt{light\_off\_attack} as many of the heatmap cells are above AUC-ROC of 0.70 (see greater proportion of redder cells placed in the heatmap). Moreover, we observe that there is a region of better performance for the \texttt{max\_engine\_attack} characterized for lower values of $\omega \le 300$ and $\delta \le 100$. Interestingly, among the highest performances across attack types is the \texttt{correlated\_attack} with an AUC-ROC of $0.77$ for the combination ($\omega=300, \delta=270$). We also observed that for the \texttt{max\_speedometer\_attack} and \texttt{light\_on\_attack} the reported AUC-ROC tends to be below $0.50$ in most of the combinations of $\omega$ and $\delta$. This suggests that the Matrix Correlation Distribution Method has a hard time detecting these attacks as the probability of ranking random positive windows more highly than random negatives windows is below $0.50$ in most of the $\omega$ and $\delta$ combinations. 
\begin{figure*}[!htp]
\centering
\includegraphics[width=1.0\textwidth]{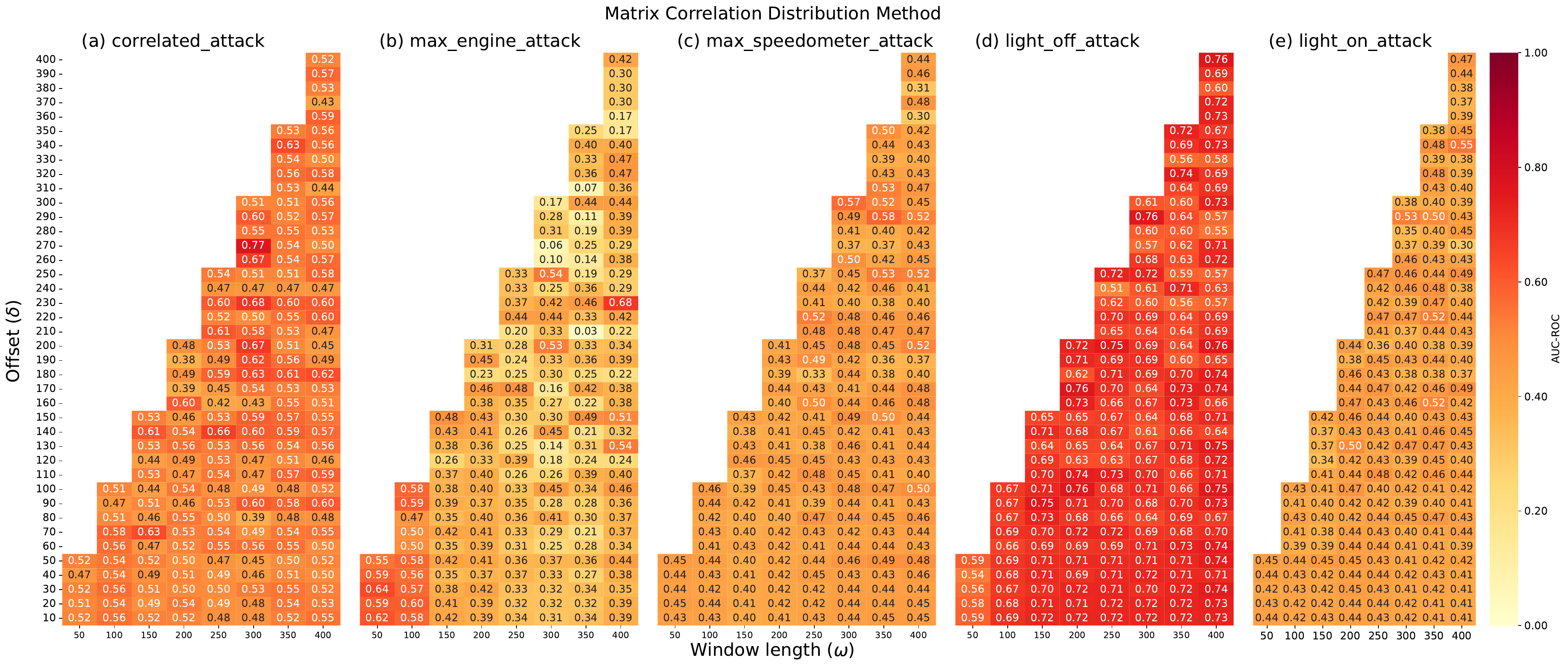}
\caption{AUC-ROC over different window length and offset combinations for the \textit{Matrix Correlation Distribution} method on the different attacks in the ROAD dataset: (a) correlated attack, (b) max engine coolant temperature attack, (c) max speedometer attack, (d) reverse light off attack, and (e) reverse light on attack.}
\label{fig:distribution method}
\end{figure*}

Fig.~\ref{fig:correlation method} shows results for the Matrix Correlation Correlation Method. We observe that the highest detection performance is obtained for the \texttt{correlated\_attack} with an AUC-ROC of $0.89$ for the combination ($\omega=150, \delta=130$). We also observe AUC-ROC values above $0.70$ for the \texttt{correlated\_attack} that tend to be generally contained at lower values of $\omega$ and $\delta$, i.e., $\omega \le 200$ and $\delta \le 150$. Note that the evaluation metrics for the remaining attacks (i.e., \texttt{max\_speedometer\_attack}, \texttt{max\_engine\_attack}, and \texttt{light\_on\_attack}) are not as good as in the previous attacks barely staying  above $0.50$. An exception is the \texttt{light\_off\_attack} where AUC-ROC values around 0.70 are contained at lower values of $\omega$ and $\delta$, i.e., $\omega \le 150$ and $\delta \le 150$. 
\begin{figure*}[!htp]
\centering
\includegraphics[width=1.0\textwidth]{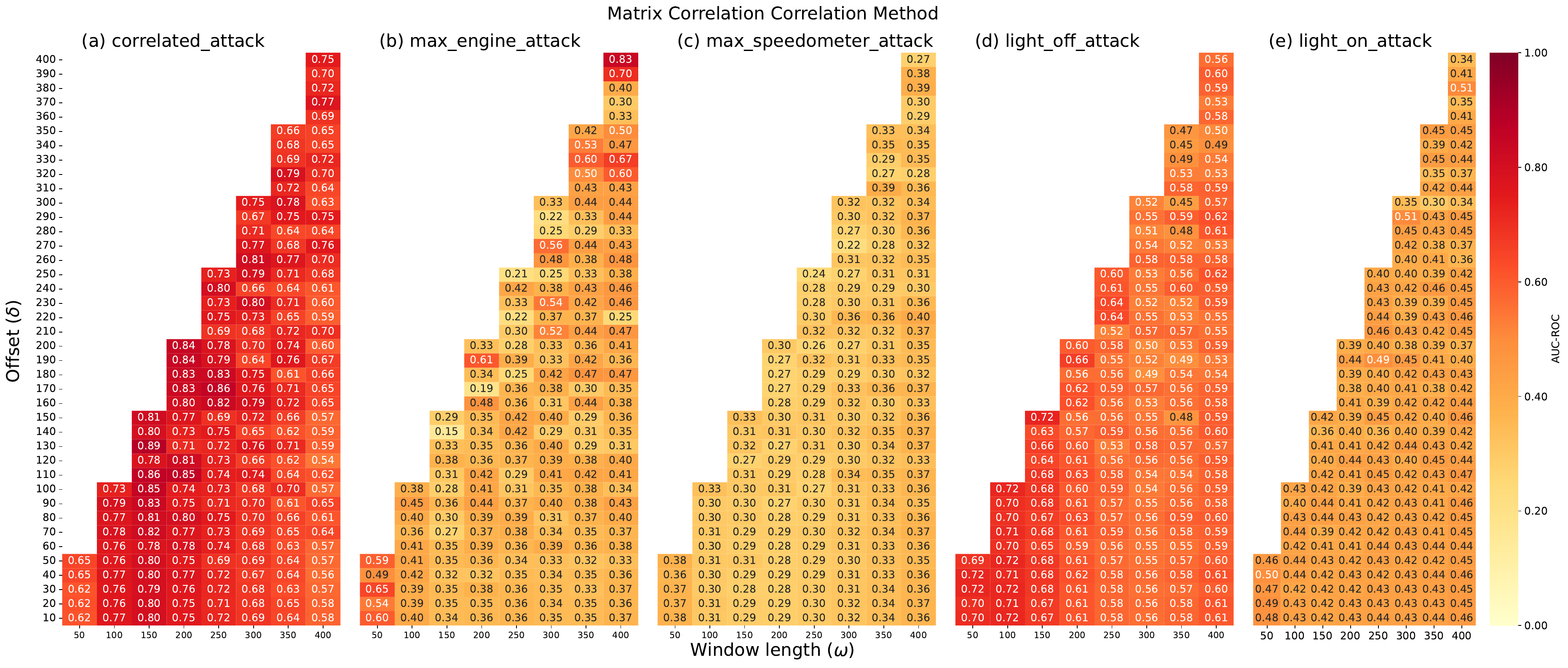}
\caption{AUC-ROC over different window length and offset combinations for the \textit{Matrix Correlation Correlation} method on the different attacks in the ROAD dataset: (a) correlated attack, (b) max engine coolant temperature attack, (c) max speedometer attack, (d) reverse light off attack, and (e) reverse light on attack.}
\label{fig:correlation method}
\end{figure*}

Fig.~\ref{fig:dbscan method} shows results for the Ganesan17 Method. We observe that the highest evaluation metric is achieved for the \texttt{max\_speedometer\_attack} at $0.74$ for the combination ($\omega=400, \delta=380$). For this attack, we also found a region where AUC-ROC rose $0.70$ specially contained at lower values of $\omega$ and $\delta$, i.e., $100 \le \omega \le 150$ and $\delta \le 160$. For the remaining attack types we did not find any other remarkable evaluation metrics. 
\begin{figure*}[!htp]
\centering
\includegraphics[width=1.0\textwidth]{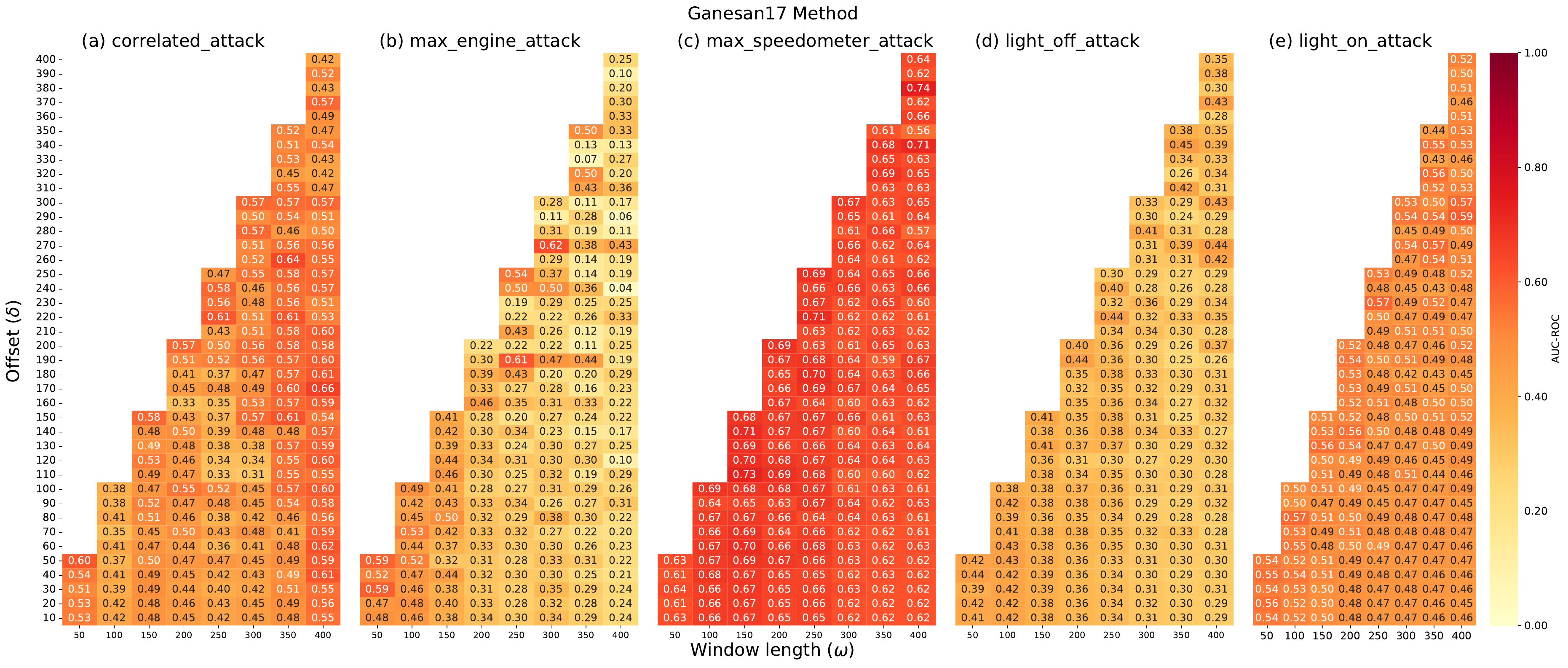}
\caption{AUC-ROC over different window length and offset combinations for the \textit{Ganesan17} method on the different attacks in the ROAD dataset: (a) correlated attack, (b) max engine coolant temperature attack, (c) max speedometer attack, (d) reverse light off attack, and (e) reverse light on attack.}
\label{fig:dbscan method}
\end{figure*}

Fig.~\ref{fig:ahc method} shows results for the Moriano22 Method. We observe that the maximum AUC-ROC among attack types is achieved with the \texttt{max\_engine\_ attack} achieving an AUC-ROC of $1.00$ for a few combinations of $\omega$ and $\delta$. Additionally, for this attack, we found a region where AUC-ROC rose $0.70$ that is contained at lower values of $\omega$ and $\delta$, i.e., $100 \le \omega \le 400$ and $10 \le \delta \le 160$. For the \texttt{light\_off\_attack}, we also observe a performance region with values of AUC-ROC above $0.70$, generally contained at lower values of $\omega$ and $\delta$, i.e., $\omega \le 200$ and $\delta \le 200$. This method also displays AUC-ROC values above $0.60$ for the \texttt{light\_on\_attack} generally contained at the entire domain of values of $\omega$ and $\delta$. In addition, we observe that the evaluation metrics for the remaining attacks (i.e., \texttt{correlated\_attack} and \texttt{max\_speedometer\_attack}) are not as good as in the previous attacks barely reaching  above $0.50$.
\begin{figure*}[!htp]
\centering
\includegraphics[width=1.0\textwidth]{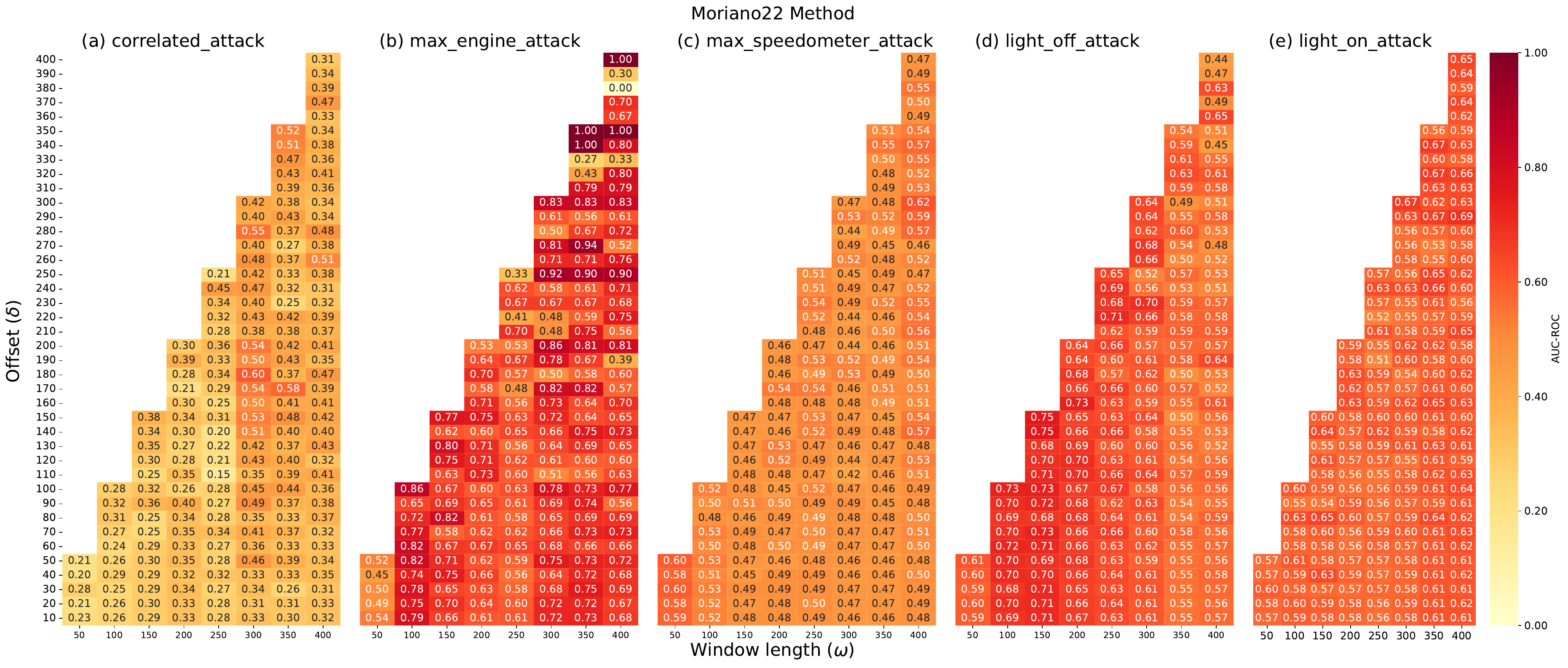}
\caption{AUC-ROC over different window length and offset combinations for the \textit{Moriano22} method on the different attacks in the ROAD dataset: (a) correlated attack, (b) max engine coolant temperature attack, (c) max speedometer attack, (d) reverse light off attack, and (e) reverse light on attack.}
\label{fig:ahc method}
\end{figure*}

\hr{To assist interpretation, we also provide Table~\ref{table: AUC-ROC summary table}, which report average trends across configurations.}

\subsection{Evaluation Metrics Summary} \label{subsec:Evaluation:Metrics:Summary}

Table~\ref{table: AUC-ROC summary table} summarizes results extracted from the heatmaps. We place attack categories in the rows and attack detection methods in the columns. This means that each cell in this table shows the performance of a specific detection method in a particular attack from the ROAD dataset. We focus on summary statistics including the mean ($\mu$), standard deviation ($\sigma$), median ($\eta$), minimum ($\min$) and maximum ($\max$). We also show the average and standard deviations from the maximum values for each attack detection methods and attack types in the ROAD dataset. They are displayed in the last row and column in the table. 

Notably, from the attack detection methods, Moriano22 Method performs the best across attacks (i.e., $\overline{max}=0.73$), followed by Matrix Correlation Distribution and Matrix Correlation Correlation (i.e., $\overline{max}=0.67$), and Ganesan17 (i.e., $\overline{max}=0.61$). Among these, the minimum and maximum variation of maximum AUC-ROC is obtained by Matrix Correlation Distribution (i.e., $\sigma_{max}=0.10$) and Matrix Correlation Correlation (i.e., $\sigma_{max}=0.21$) respectively. 

Note that, from the attack categories, the \texttt{max\_engine\_attack} was found to be the attack that is detected with higher AUC-ROC among bechmarked algorithms (i.e., $\overline{max}=0.79$), followed by \texttt{correlated\_attack} (i.e., $\overline{max}=0.73$), \texttt{light\_off\_attack} (i.e., $\overline{max}=0.67$) and \texttt{max\_speedometer\_attack} and \texttt{light\_on\_attack} (i.e., (i.e., $\overline{max}=0.59$). Among these, the minimum and maximum variation of maximum AUC-ROC is obtained by \texttt{light\_on\_attack} (i.e., $\sigma_{max}=0.00$) and \texttt{max\_engine\_attack} (i.e., $\sigma_{max}=0.17$) respectively. 

\begin{table*}[htb]
\caption{Summary of evaluation metrics based on the AUC-ROC heatmaps. We show the mean ($\mu$), standard deviation ($\sigma$), median ($\eta$), minimum ($\min$) including the combination of window length and offset that produces it, and maximum ($\max$) including the combination of window length and offset that produces it. Attack types are placed in the rows while detection methods are placed in the columns. We also report averages and standard deviation of the maximum values obtained by each attack and detection method denoted by $\overline{max}$ and $\sigma_{max}$.}
\begin{center}
\footnotesize
\begin{tabular}{|p{1.2in}|l|l|l|l|l l|}
    \hline
    {\centering Attack/Method} & {\centering Matrix Correlation} & {\centering Matrix Correlation} & {\centering Ganesan17} & {\centering Moriano22} & & \\
    & {\centering Distribution} & {\centering Correlation} & & & & \\
    \hline
    \multirow{6}{*}{correlated\_attack} & $\mu=0.53$ & $\mu=0.71$ & $\mu=0.49$ & $\mu=0.35$ & \multirow{5}{*}{\rotatebox[origin=r]{90}{$\overline{max}= 0.73$}} & \multirow{5}{*}{\rotatebox[origin=r]{90}{$\sigma_{max}= 0.12$}}  \\
    & $\sigma=0.06$ & $\sigma=0.07$ & $\sigma=0.07$ & $\sigma=0.08$ & & \\
    & $\eta=0.53$ & $\eta=0.72$ & $\eta=0.49$ & $\eta=0.34$ & & \\
    & $\min=0.38$ (200, 190) & $\min=0.54$ (400, 120) & $\min=0.31$ (300, 110) & $\min=0.15$ (250, 110) & & \\
    & $\max=0.77$ (300, 270) & $\max=0.88$ (150, 130) & $\max=0.66$ (400, 170) & $\max=0.60$ (300, 180) & & \\
    & & & & & & \\
    \hline
    \multirow{6}{*}{max\_engine\_attack} & $\mu=0.36$ & $\mu=0.38$ & $\mu=0.31$ & $\mu=0.66$ & \multirow{5}{*}{\rotatebox[origin=r]{90}{$\overline{max}= 0.79$}} & \multirow{5}{*}{\rotatebox[origin=r]{90}{$\sigma_{max}= 0.17$}} \\
    & $\sigma=0.11$ & $\sigma=0.09$ & $\sigma=0.11$ & $\sigma=0.13$ & & \\
    & $\eta=0.36$ & $\eta=0.37$ & $\eta=0.29$ & $\eta=0.67$ & & \\
    & $\min=0.03$ (350, 210) & $\min=0.15$ (150, 140) & $\min=0.04$ (400, 240) & $\min=0.00$ (400, 380) & & \\
    & $\max=0.68$ (400, 230) & $\max=0.83$ (400, 400) & $\max=0.63$ (300, 270) & $\max=1.00$ (400, 400) & & \\
    & & & & & & \\
    \hline
    \multirow{5}{*}{max\_speedometer\_attack} & $\mu=0.44$ & $\mu=0.32$ & $\mu=0.64$ & $\mu=0.49$ & \multirow{5}{*}{\rotatebox[origin=r]{90}{$\overline{max}= 0.59$}} & \multirow{5}{*}{\rotatebox[origin=r]{90}{$\sigma_{max}= 0.14$}} \\
    & $\sigma=0.04$ & $\sigma=0.03$ & $\sigma=0.03$ & $\sigma=0.03$ & & \\
    & $\eta=0.43$ & $\eta=0.31$ & $\eta=0.64$ & $\eta=0.49$ & & \\
    & $\min=0.30$ (400, 360) & $\min=0.22$ (300, 270) & $\min=0.56$ (400, 350) & $\min=0.42$ (300, 100) & & \\
    & $\max=0.58$ (350, 290) & $\max=0.40$ (400, 220) & $\max=0.74$ (400, 380) & $\max=0.62$ (400, 300) & & \\
    & & & & & & \\
    \hline
    \multirow{6}{*}{light\_off\_attack} & $\mu=0.68$ & $\mu=0.59$ & $\mu=0.34$ & $\mu=0.61$ & \multirow{5}{*}{\rotatebox[origin=r]{90}{$\overline{max}= 0.67$}} & \multirow{5}{*}{\rotatebox[origin=r]{90}{$\sigma_{max}= 0.15$}} \\
    & $\sigma=0.05$ & $\sigma=0.06$ & $\sigma=0.05$ & $\sigma=0.06$ & & \\
    & $\eta=0.69$ & $\eta=0.58$ & $\eta=0.34$ & $\eta=0.61$ & & \\
    & $\min=0.51$ (250, 240) & $\min=0.45$ (350, 300) & $\min=0.24$ (350, 290) & $\min=0.44$ (400, 400) & & \\
    & $\max=0.76$ (400, 400) & $\max=0.72$ (150, 150) & $\max=0.45$ (350, 340) & $\max=0.75$ (150, 140) & & \\
    & & & & & & \\
    \hline
    \multirow{5}{*}{light\_on\_attack} & $\mu=0.42$ & $\mu=0.42$ & $\mu=0.49$ & $\mu=0.59$ & \multirow{5}{*}{\rotatebox[origin=r]{90}{$\overline{max}= 0.59$}} & \multirow{5}{*}{\rotatebox[origin=r]{90}{$\sigma_{max}= 0.08$}} \\
    & $\sigma=0.03$ & $\sigma=0.03$ & $\sigma=0.03$ & $\sigma=0.03$ & & \\
    & $\eta=0.43$ & $\eta=0.42$ & $\eta=0.49$ & $\eta=0.59$ & & \\
    & $\min=0.29$ (400, 270) & $\min=0.28$ (350, 300) & $\min=0.42$ (300, 180) & $\min=0.51$ (250, 190) & & \\
    & $\max=0.55$ (400, 340) & $\max=0.51$ (300, 290) & $\max=0.59$ (400, 290) & $\max=0.69$ (400, 290) & & \\
    & & & & & & \\
    \hline
    & $\overline{max} = 0.67$ & $\overline{max} = 0.67$ & $\overline{max} = 0.61$ & $\overline{max} = 0.73$ & & \\
    & $\sigma_{max} = 0.10$ & 
    $\sigma_{max} = 0.21$ & 
    $\sigma_{max} = 0.11$ &
    $\sigma_{max} = 0.16$ & & \\
    \hline
\end{tabular}
\end{center}
\label{table: AUC-ROC summary table}
\end{table*}

Additionally, Fig.~\ref{fig:polar plot} compares each detection method on the basis of summary statistics for each of attack type using dot plots. A desirable online algorithm for detecting masquerade attacks in CAN by processing streams of time series data should have high values of $\mu$, $\eta$, $\min$, and $\max$ and lower values of $\sigma$ in the range $[0, 1]$. We notice that there is no free lunch in the detection of masquerade attacks in CAN in an online setting as there is no single algorithm that is optimal for detecting each attack type. Notably, Ganesan17, the worst performing algorithms across all attack types is the best performing detection algorithm for the hardest attack to detect, i.e., \texttt{max\_speedometer\_attack}. 
\begin{figure*}[!htp]
    \centering
    \includegraphics[width=\linewidth]{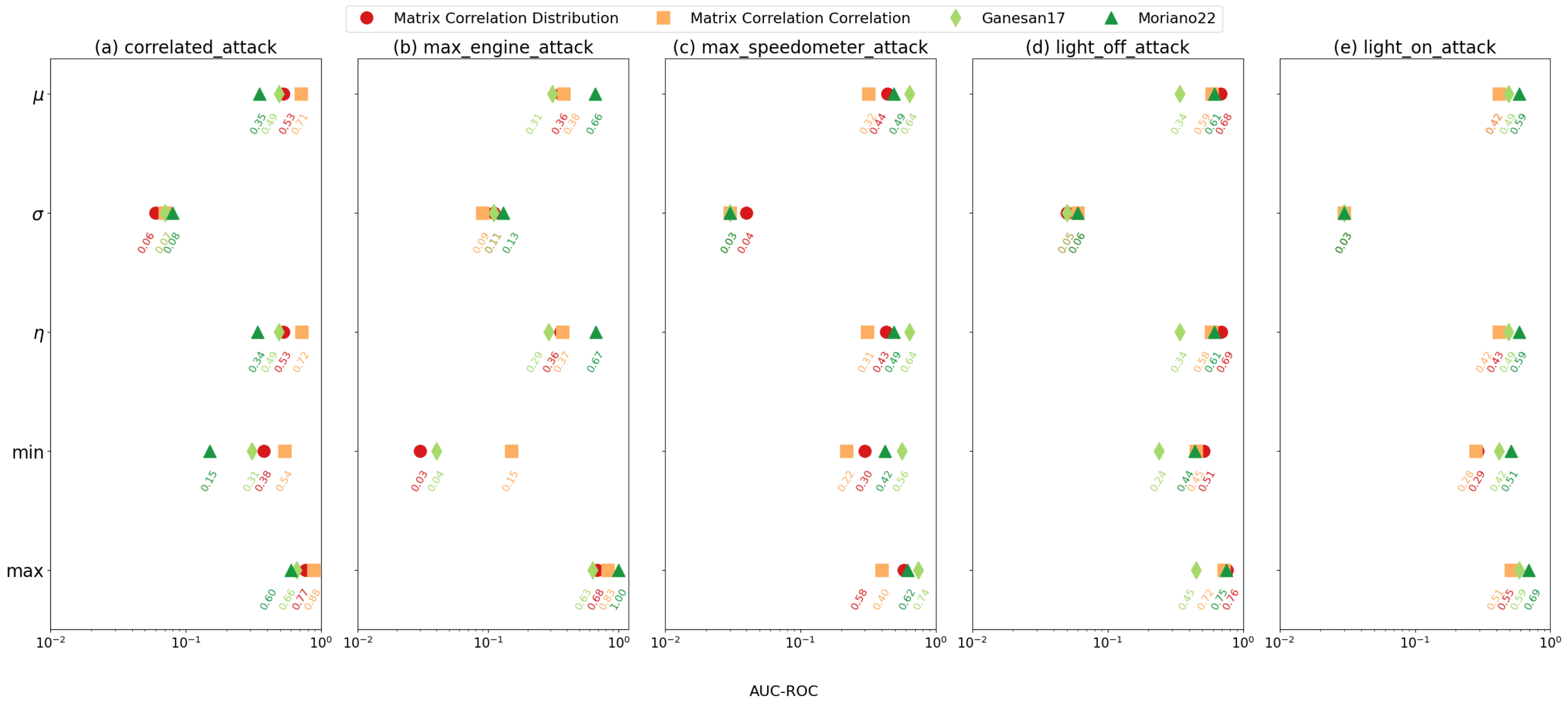}
    \caption{Dot plots comparing each detection method based on summary statistics of AUC-ROC per attack type: (a) \texttt{correlated\_attack}, (b) \texttt{max\_engine\_attack}, (c) \texttt{max\_speedometer\_attack}, (d) \texttt{light\_off\_attack}, and (e) \texttt{light\_on\_attack}.}
    \label{fig:polar plot}
\end{figure*}

Finally, we performed grid search over predefined hyperparameter ranges for Moriano22. Specifically, we varied $r \in \{-5, -3, \ldots, 3 \}$ and $\alpha \in \{0.1, 0.2, \dots, 0.8\}$. These ranges cover a variety of standard configurations. To ensure a thorough evaluation, we applied this procedure to the specific combination of $\omega$ and $\delta$, where Moriano22 performed the worst among all attack scenarios. This allowed us to analyze the worst-case scenario and assess the performance improvements achievable by tuning hyperparameters. We visualized the sensitivity of AUC-ROC scores to these parameters using heatmaps. For each attack category, we computed the average performance and summarized it in a single heatmap. This approach helps interpret the effect of hyperparameter tuning on the method's performance. Figure~\ref{fig: Moriano22:Hyperparameters} illustrates how AUC-ROC scores improve with optimized combinations of $r$ and $\alpha$. Notably, scores increased significantly from their minimum values (see Table~2 column Moriano22) to their maximum values shown in the heatmaps, except for the \texttt{light\_off\_attack} category. In addition, Table~\ref{table: AUC-ROC hyperparameter} compares the AUC-ROC values of the Moriano22 method before and after hyperparameter optimization. This analysis confirms the robustness of the proposed method under optimized settings while highlighting areas for further optimization.
\begin{figure*}[!htp] 
\centering
\includegraphics[width=1.0\linewidth]{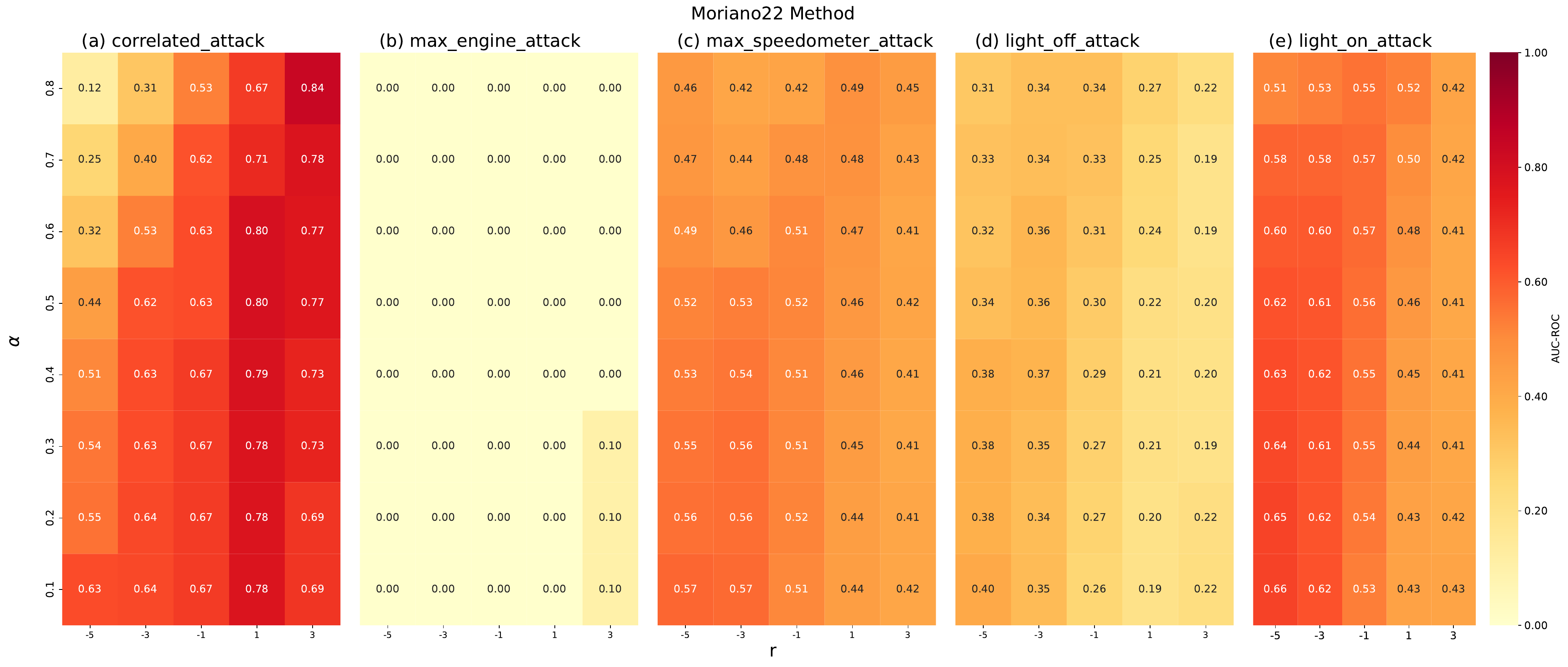} 
\caption{AUC-ROC values for Moriano22 across different combinations of $r$ and $\alpha$ in the ROAD dataset attacks: (a) correlated attack, (b) max engine coolant temperature attack, (c) max speedometer attack, (d) reverse light off attack, and (e) reverse light on attack.}
\label{fig: Moriano22:Hyperparameters}
\end{figure*}

\begin{table*}[htb]
\caption{Comparison of AUC-ROC values for the Moriano22 method before and after hyperparameter optimization. The ``before'' column shows values from Table~2, where the method produces the lowest result. The ``after'' column presents values obtained using the window length and offset combination from the ``before'' column. The corresponding $r$ and $\alpha$ combinations that achieve the maximum values are shown in parentheses. Attack types are listed in the rows, while results before and after optimization are in the columns. The last column reports the absolute percentage change.}
\begin{center}
\footnotesize
\begin{tabular}{|p{1.5in}|l|l|c|}
    \hline
    {\centering Attack/Method} & {\centering Moriano22 Before} & {\centering Moriano22 After} & Absolute \\
    & {\centering Optimization} & {\centering  Optimization} & Change (\%)  \\
    \hline 
    \texttt{correlated\_attack} & 0.15 & 0.84 (3, 0.8) & +69\% \\
    & & & \\
    \hline
    \texttt{max\_engine\_attack} & 0.00 & 0.10 (3, 0.3) & +10\%  \\
    & & & \\
    \hline
    \texttt{max\_speedometer\_attack} & 0.42 & 0.57 (-5, 0.1) & +15\%  \\
    & & & \\
    \hline
    \texttt{light\_off\_attack} & 0.44 & 0.40 (-5, 0.1) & -4\%  \\ 
    & & & \\
    \hline
    \texttt{light\_on\_attack} & 0.51 & 0.66 (-5, 0.1) &  15\% \\
    & & & \\
    \hline
\end{tabular}
\end{center}
\label{table: AUC-ROC hyperparameter}
\end{table*}

\subsection{Performance Metrics Summary} \label{subsec:Perfromance:Metrics:Summary}

Table~\ref{table: TTW summary table} summarizes TTW results based on heatmaps that we do not display here. We place attack categories in the rows and attack detection methods in the columns. This means that each cell in this table shows the TTW of a specific detection method
in a particular attack from the ROAD dataset. We focus on summary statistics including the mean ($\mu$), standard deviation ($\sigma$), median ($\eta$), minimum ($\min$), and maximum ($\max$). In addition, we also show the average and standard deviations from the minimum values for each attack detection methods and attack types in the ROAD dataset. They are displayed as the last row and column in the
table. 

Notably, from the attack detection methods, Matrix Correlation Distribution is the quickest across attacks (i.e., $\overline{min}$ = 0.33), followed by Matrix Correlation Correlation (i.e., $\overline{min}$ = 2.53), Ganesan17 (i.e., $\overline{min}$ = 3.65), and Moriano22 (i.e., $\overline{min}$ = 8.42). Among these, the minimum and maximum variation of maximum TTW is obtained
by Matrix Correlation Distribution (i.e., $\sigma_{min}$ = 0.00) and
Moriano22 (i.e., $\sigma_{min}$ = 0.51) respectively.

In addition, from the attack categories, we found that there are no outstanding variations on the TTW within attack categories varying from 3.58 (i.e., \texttt{light\_off\_attack}) to 3.89 (\texttt{light\_on\_attack}) on average.

Finally, Fig.~\ref{fig:polar plot min} compares each detection method on the basis of TTW summary statistics for each attack type using dot plots. A desirable online algorithm for detecting masquerade attacks in CAN by processing streams of time series data should have low values of $\mu$, $\eta$, $\min$, $\max$ and lower values of $\sigma$ in the range $[0, 1]$. Remarkably, Matrix Correlation Distribution method is the quickest detection algorithm one order of magnitude lower than the others. Note that, the benchmarked algorithms infer in near real-time, with TTW values typically within a few milliseconds. Since human driver response times range from 0.7 to 1.5 seconds~\citep{Drozdziel:2020:Drivers:Reaction:Time:Real:Traffic}, the algorithms meet the requirement of inferring well below the lower bound of 0.7 seconds.

\begin{figure*}[!htp]
    \centering
    \includegraphics[width=\linewidth]{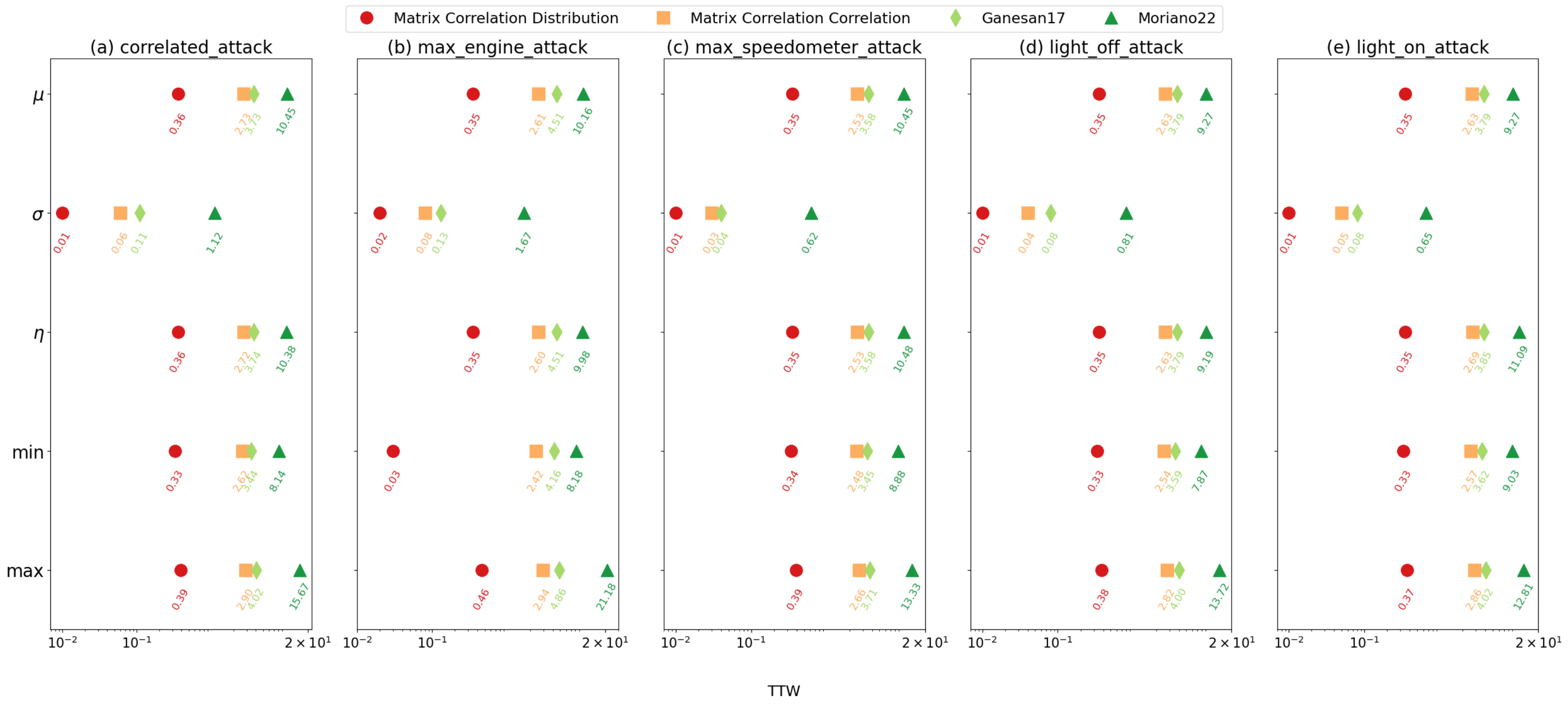}
    \caption{Dot plots comparing each detection method based on summary statistics of TTW per attack type: (a) \texttt{correlated\_attack}, (b) \texttt{max\_engine\_attack}, (c) \texttt{max\_speedometer\_attack}, (d) \texttt{light\_off\_attack}, and (e) \texttt{light\_on\_attack}.}
    \label{fig:polar plot min}
\end{figure*}

\begin{table*}[htb]
\caption{Summary of performance metrics based on TTW in miliseconds. We show the mean ($\mu$), standard deviation ($\sigma$), median ($\eta$), minimum ($\min$) including the combination of window length and offset that produces it, and maximum ($\max$) including the combination of window length and offset that produces it. Attack types are placed in the rows while detection methods are placed in the columns. We also report averages and standard deviation of the minimum values obtained by each attack and detection method denoted by $\overline{min}$ and $\sigma_{min}$.}
\begin{center}
\footnotesize
\begin{tabular}{|p{1.2in}|l|l|l|l|l l|}
    \hline
    {\centering Attack/Method} & {\centering Matrix Correlation} & {\centering Matrix Correlation} & {\centering Ganesan17} & {\centering Moriano22} & & \\
    & {\centering Distribution} & {\centering Correlation} & & & & \\
    \hline
    \multirow{6}{*}{correlated\_attack} & $\mu=0.36$ & $\mu=2.73$ & $\mu=3.73$ & $\mu=10.45$ & \multirow{5}{*}{\rotatebox[origin=r]{90}{$\overline{min}= 3.63$}} & \multirow{5}{*}{\rotatebox[origin=r]{90}{$\sigma_{min}= 3.28$}}  \\
    & $\sigma=0.01$ & $\sigma=0.06$ & $\sigma=0.11$ & $\sigma=1.12$ & & \\
    & $\eta=0.36$ & $\eta=2.72$ & $\eta=3.74$ & $\eta=10.38$ & & \\
    & $\min=0.33$ (50, 40) & $\min=2.62$ (100, 30) & $\min=3.44$ (250, 240) & $\min=8.14$ (50, 20) & & \\
    & $\max=0.39$ (400, 320) & $\max=2.90$ (250, 140) & $\max=4.02$ (350, 250) & $\max=15.67$ (300, 160) & & \\
    & & & & & & \\
    \hline
    \multirow{6}{*}{max\_engine\_attack} & $\mu=0.35$ & $\mu=2.61$ & $\mu=4.51$ & $\mu=10.16$ & \multirow{5}{*}{\rotatebox[origin=r]{90}{$\overline{min}= 3.77$}} & \multirow{5}{*}{\rotatebox[origin=r]{90}{$\sigma_{min}= 3.33$}} \\
    & $\sigma=0.02$ & $\sigma=0.08$ & $\sigma=0.13$ & $\sigma=1.67$ & & \\
    & $\eta=0.35$ & $\eta=2.60$ & $\eta=4.51$ & $\eta=9.98$ & & \\
    & $\min=0.03$ (150, 10) & $\min=2.42$ (100, 60) & $\min=4.16$ (50, 50) & $\min=8.18$ (50, 40) & & \\
    & $\max=0.46$ (150, 140) & $\max=2.94$ (400, 360) & $\max=4.86$ (400, 270) & $\max=21.18$ (350, 3300) & & \\
    & & & & & & \\
    \hline
    \multirow{5}{*}{max\_speedometer\_attack} & $\mu=0.35$ & $\mu=2.53$ & $\mu=3.58$ & $\mu=10.45$ & \multirow{5}{*}{\rotatebox[origin=r]{90}{$\overline{min}= 3.79$}} & \multirow{5}{*}{\rotatebox[origin=r]{90}{$\sigma_{min}= 3.64$}} \\
    & $\sigma=0.01$ & $\sigma=0.03$ & $\sigma=0.04$ & $\sigma=0.62$ & & \\
    & $\eta=0.35$ & $\eta=2.53$ & $\eta=3.58$ & $\eta=10.48$ & & \\
    & $\min=0.34$ (50, 20) & $\min=2.48$ (250, 240) & $\min=3.45$ (300, 270) & $\min=8.88$ (50, 40) & & \\
    & $\max=0.39$ (100, 90) & $\max=2.66$ (100, 80) & $\max=3.71$ (100, 80) & $\max=13.33$ (400, 340) & & \\
    & & & & & & \\
    \hline
    \multirow{6}{*}{light\_off\_attack} & $\mu=0.35$ & $\mu=2.63$ & $\mu=3.79$ & $\mu=9.27$ & \multirow{5}{*}{\rotatebox[origin=r]{90}{$\overline{min}= 3.58$}} & \multirow{5}{*}{\rotatebox[origin=r]{90}{$\sigma_{min}= 3.16$}} \\
    & $\sigma=0.01$ & $\sigma=0.04$ & $\sigma=0.08$ & $\sigma=0.81$ & & \\
    & $\eta=0.35$ & $\eta=2.63$ & $\eta=3.79$ & $\eta=9.19$ & & \\
    & $\min=0.33$ (50, 30) & $\min=2.54$ (100, 60) & $\min=3.59$ (350, 340) & $\min=7.87$ (50, 60) & & \\
    & $\max=0.38$ (300, 220) & $\max=2.82$ (400, 350) & $\max=4.00$ (400, 200) & $\max=13.72$ (400, 260) & & \\
    & & & & & & \\
    \hline
    \multirow{5}{*}{light\_on\_attack} & $\mu=0.35$ & $\mu=2.63$ & $\mu=3.79$ & $\mu=9.27$ & \multirow{5}{*}{\rotatebox[origin=r]{90}{$\overline{min}= 3.89$}} & \multirow{5}{*}{\rotatebox[origin=r]{90}{$\sigma_{min}= 3.69$}} \\
    & $\sigma=0.01$ & $\sigma=0.05$ & $\sigma=0.08$ & $\sigma=0.65$ & & \\
    & $\eta=0.35$ & $\eta=2.69$ & $\eta=3.85$ & $\eta=11.09$ & & \\
    & $\min=0.33$ (50, 40) & $\min=2.57$ (400, 380) & $\min=3.62$ (100, 90) & $\min=9.03$ (50, 40) & & \\
    & $\max=0.37$ (400, 180) & $\max=2.86$ (200, 200) & $\max=4.02$ (350, 290) & $\max=12.81$ (400, 320) & & \\
    & & & & & & \\
    \hline
    & $\overline{min} = 0.33$ & $\overline{min} = 2.53$ & $\overline{min} = 3.65$ & $\overline{min} = 8.42$ & & \\
    & $\sigma_{min} = 0.00$ & 
    $\sigma_{min} = 0.08$ & 
    $\sigma_{min} = 0.30$ &
    $\sigma_{min} = 0.51$ & & \\
    \hline
\end{tabular}
\end{center}
\label{table: TTW summary table}
\end{table*}

\section{Discussion} \label{sec:Discussion}

In this work, we have comparative evaluated non-DL-based unsupervised online IDS for detecting masquerade attacks in CAN data. Our results indicate that the evaluated algorithms are not equally effective at detecting every attack type as it has been shown before in previous studies in the offline setting~\citep{Agbaje:2022:Framework:CAN:IDS:Evaluation, Sharmin:2023:Benchmark:CAN:IDS:ROAD, Pollicino:2023:Performance:Timing:IDS:Comparison}. We attributed this to the foundational principle used for each of the detection algorithms and the magnitude of the perturbation that is incurred when introducing each of the attacks. Overall, Moriano22 produces the best detection results across attack types. Among attacks, \texttt{max\_speedomer\_attack} and \texttt{light\_on\_attack} are the overall hardest to detect attack type. 

In agreement with the foundational principle of Moriano22 (i.e., finding disruptions in the hierarchy of time series clusters), our results supports the fact that introduced masquerade attacks in the ROAD dataset change the relative hierarchy of time series clusters as it was shown previously in an offline setting in~\citep{Moriano:2022:AHC}. As opposed to Ganesan17, which focuses on significant deviations of time series cluster assignations, Moriano22 captures significant deviations in the hierarchy of clusters. Contrary to our intuition, our findings also reveal that Matrix Correlation Distribution and Matrix Correlation Correlation methods perform better than Ganesan17. We suspect that this is in part because of the multiple hyperparameters that need to be adjusted in Ganesan17 to tune its optimal performance. 

Although Moriano22 outperformed the other methods, achieving the best detection results across attack types, it showed lower average AUC-ROC values for the \texttt{correlated\_attack} (i.e., $\mu$ = 0.35). This result likely stems from the windowing approach used in the testing, which limits capturing longer-term effects of masquerade attacks. These longer-term effects may hinder the formation of consistent cluster hierarchies and could be further explored with other time series prequential evaluation methods~\citep{Gama:2014:Survey:Concept:Drift}. However, Moriano22 outperformed the other methods on the remaining attacks, i.e.,\texttt{max\_engine\_attack}, \texttt{max\_speedometer\_attack}, \texttt{light\_off\_attack}, and \texttt{light\_on\_attack}. This indicates greater robustness in detecting masquerade attacks, particularly in the presence of subtle data manipulations such as those affecting the engine coolant, speedometer, and reverse light signals. Given the variability in performance across attacks, we expect that combining the comparative evaluationed algorithms through ensemble learning could improve overall robustness. An ensemble could exploit the strengths of individual methods and compensate for their weaknesses, especially when detecting subtle or sustained attacks.

While previous research has focused on detecting masquerade attacks in an offline setting, our results demonstrate the importance of considering how sliding windows are processed in an online setting with respect to window length and offset. Specifically, our results provide new insights on the effect of choosing carefully these two data partitioning parameters. To make comparative evaluationed methods more effective to work in a real-world setting, online algorithms should strive to perform well when processing small batches of data (i.e., lower $\omega$ values) and low latency between window transitions (i.e., lower $\delta$ values). We quantify these relationships for every method on each of the attack types.

\hr{While we recognize the value of deep learning-based IDSs, especially those validated on preselected signals or simplified environments, our comparative evaluation deliberately focuses on unsupervised non-DL methods that can operate on high-dimensional, unlabeled signal-level data without requiring prior semantic knowledge. Given the variability and scale of the ROAD dataset, along with the challenges in training, tuning, and deploying DL models in production vehicular environments, we restrict our analysis to more transparent and resource-efficient methods. Our released code provides an open pathway for future researchers to incorporate and evaluate DL-based methods under identical streaming configurations.}

\hr{We acknowledge several limitations of our study, which we elaborate in the following section.}

\section{Threats to Validity} \label{sec:Threats to Validity}

\hr{Our study provides insights on the effectiveness and performance of online detection algorithms for detecting masquerade attacks from the ROAD dataset. Having noted the strengths of our comparative evaluation framework, we are also aware of the following limitations of our proposed work.}

\hr{\noindent \textbf{Lack of Validation in a Real Environment:} Our proposed framework partitions logs of already collected CAN captures to simulate an streaming environment of CAN data using windows. We did not empirically validated our results on a moving vehicle including the necessary hardware and performance tuning involved to do so. \hr{We acknowledge that true in-vehicle IDS deployment would require additional validation on automotive-grade microcontrollers. Our goal here is to evaluate algorithmic performance under realistic timing constraints using resource-constrained proxies, which is a common practice in CAN IDS literature.}}

\hr{\noindent \textbf{Limited Number of Comparative Evaluationed Algorithms:} We evaluated four unsupervised online detection algorithms. As we strive for computationally efficient approaches and test their efficacy on a widely known attack dataset, we did not include unsupervised DL-based approaches (e.g., \citep{Hanselmann:2020:CANet, Shahriar:2023:CANShield}) in the comparison.}

\hr{\noindent \textbf{Focus on a Single Testbed Dataset:} We used the ROAD dataset with a limited number of masquerade attacks to comparative evaluation online detection algorithms. We did not consider newer datasets with a higher number of masquerade attacks generated on modern automobiles~\citep{Rajapaksha:2024:CAN-MIRGU}. While we focus on the ROAD dataset due to its broad signal coverage (hundreds of IDs and signals) and real driving instrumentation, we acknowledge that other datasets like CrySyS (Gazdag et al.~\citep{Gazdag:2023:Crysys:Dataset}) offer complementary perspectives. CrySyS introduces signal modification on up to two signals in both physical and simulated environments, and is a valuable resource for future cross-dataset evaluation. Our modular comparative evaluation infrastructure and public code allow straightforward extension to include CrySyS or similar datasets. We consider this a promising avenue for subsequent work.}

\hr{\noindent \textbf{Use of Default Parameter Values in Detection Algorithms:} We used default parameters values as the original algorithms in Ganesan17 and Moriano22. We optimized the hyperparameters for Moriano22, focusing on a single combination of $\omega$ and $\delta$ that achieved the lowest AUC-ROC with default settings. This approach confirms the method's robustness under optimized conditions and identifies areas for further improvement.}

\hr{We anticipate that field experiments on a moving vehicle can be performed to validate our results based on new technologies being deployed to extract time series data from CAN payloads (e.g., \citep{Verma:2021:CAN-D}).}

\section{Conclusion}
This paper presents a comparative evaluation of non-DL-based unsupervised online IDS for detecting masquerade attacks in CAN. We focus on four detection algorithms (i.e., Matrix Correlation Distribution, Matrix Correlation Correlation, Ganesan17~\citep{Ganesan:2017:Exploiting:Correlations:Heterogeneous:Sensors}, and Moriano22~\citep{Moriano:2022:AHC}) which foundational mechanism is based on capturing significant deviations from expected pairwise correlations over a stream of multivariate time series. The inclusion of these baseline algorithms along with their open source implementation allow direct comparison with forthcoming online algorithms in this research space. The presented evaluation differs from other evaluations in prior works in that it focuses on the online operation of detection algorithms (as opposed to offline operation mode) aimed at detecting masquerade attacks, i.e., the stealthiest category of attacks in the CAN bus. Our evaluation is based on the concept of time windows that control the partition of a stream of time series. Specifically, we use sliding windows, with respect to the number of observations, to analyze variations in the AUC-ROC based on different combinations of window length and offset. We report conditions that makes evaluated algorithms more successful than others for certain attack types as well a summary of their overall performance on different attacks. We show that overall, Moriano22 performs the best at detecting the different attack types in the ROAD dataset. Our analysis reveals that the \texttt{max\_speedometer\_attack} is the more elusive to detect among detection algorithms.

Future work in this area involve the inclusion of more recent online detection algorithms algorithms such as the work by Shahriar et al.~\citep{Shahriar:2023:Cantropy} and the validation of our results in an empirical moving vehicle setting.

\section*{Acknowledgment} \label{sec:Acknowledgment}
This manuscript has been authored by UT-Battelle, LLC under Contract No. DE-AC05-00OR22725 with the U.S. Department of Energy. The publisher, by accepting the article for publication, acknowledges that the U.S. Government retains a non-exclusive, paid up, irrevocable, world-wide license to publish or reproduce the published form of the manuscript, or allow others to do so, for U.S. Government purposes. The DOE will provide public access to these results in accordance with the DOE Public Access Plan (\url{http://energy.gov/downloads/doe-public-access-plan}). This research was sponsored in part by Oak Ridge National Laboratory’s (ORNL’s) Laboratory Directed Research and Development program and by the DOE. There was no additional external funding received for this study. This research used birthright cloud resources of the Compute and Data Environment for Science (CADES) at the Oak Ridge National Laboratory, which is supported by the Office of Science of the U.S. Department of Energy under Contract No. DE-AC05-00OR22725. The funders had no role in study design, data collection and analysis, decision to publish, or preparation of this manuscript.

\bibliographystyle{elsarticle-num}
\bibliography{3-new_paper} 

\end{document}